\documentclass[preprint,3p]{elsarticle}

\usepackage{multirow}
\usepackage{url}
\usepackage{color}
\usepackage{hyperref}
\def    \be             {\begin{equation}}
\def    \ee             {\end{equation}}
\def    \beq             {\begin{equation}}
\def    \eeq             {\end{equation}}
\def    \ba             {\begin{eqnarray}}
\def    \ea             {\end{eqnarray}}
\def    \beqn           {\begin{eqnarray}}
\def    \eeqn           {\end{eqnarray}}
\def    \as             {\ifmmode \alpha_s \else $\alpha_s$ \fi}
\def    \nn             {\nonumber}
\def    \=              {\;=\;}
\def    \pl     #1#2#3{{Phys. Lett.} {\bf #1} (19#2) #3}
\def    \comment   #1     {{\bf *** #1 ***}}

\begin{document}

\title{Top-pair production at hadron colliders with \\
	next-to-next-to-leading logarithmic soft-gluon resummation}

\author[Paris,paris2]{Matteo Cacciari}
\author[Aachen]{Micha\l{}  Czakon}
\author[CERN]{Michelangelo Mangano}
\author[CERN]{Alexander Mitov}
\author[CERN,Milan]{Paolo Nason}

\address[Paris]{LPTHE, UPMC Univ. Paris 6 and CNRS UMR 7589, F-75252 Paris Cedex 05, France}
\address[paris2]{Universit\'e Paris Diderot, France}
\address[Aachen]{Institut f\"ur Theoretische Teilchenphysik und Kosmologie,
RWTH Aachen University, D-52056 Aachen, Germany}
\address[CERN]{Theory Division, CERN, CH-1211 Geneva 23, Switzerland}
\address[Milan]{INFN and Dept. of Physics, University of Milano Bicocca, I-20133 Milan, Italy}

\date{\today}

\cortext[thanks]{Preprint numbers: CERN-PH-TH/2011-277, TTK-11-54}

\begin{abstract}
Incorporating all recent theoretical advances, we resum soft-gluon
corrections to the total $t\bar t$ cross-section at hadron colliders
at the next-to-next-to-leading logarithmic (NNLL) order. We perform the
resummation in the well established framework of Mellin $N$-space
resummation. We exhaustively study the sources of systematic
uncertainty like renormalization and factorization scale variation,
power suppressed effects and missing two- and higher-loop
corrections. The inclusion of soft-gluon resummation at NNLL brings
only a minor decrease in the perturbative uncertainty with respect
to the NLL approximation, and a small shift in the central value,
consistent with the quoted uncertainties. These numerical
predictions agree with the currently available measurements from
the Tevatron and LHC and have uncertainty of similar size. We conclude
that significant improvements in the $t\bar t$ cross-sections can
potentially be expected only upon inclusion of the complete NNLO
corrections.
\end{abstract}

\maketitle

\section{Introduction}

The production of $t{\bar t}$ pairs at hadron colliders is well
understood within next-to-leading order (NLO) perturbative QCD, where
corrections of order ${\cal O}(\alpha_s^3)$ are included. Results have
been available for a while for the fully inclusive
\cite{Nason:1987xz,Beenakker:1988bq,Czakon:2008ii}, one-particle
inclusive \cite{Nason:1989zy,Beenakker:1990maa}, two-particle inclusive
production \cite{Mangano:1991jk}, including decay
\cite{Melnikov:2009dn}, spin correlations
\cite{Bernreuther:2001bx,Bernreuther:2004jv}, off-shell effects \cite{Denner:2010jp,Bevilacqua:2010qb} and associated jet production
\cite{Dittmaier:2007wz,Dittmaier:2008uj,Bredenstein:2009aj,Bredenstein:2010rs,Bevilacqua:2009zn,Bevilacqua:2010ve,Bevilacqua:2011hy}. In addition, the resummation
of next-to-leading logarithmic (NLL) soft gluon
effects has been long established~\cite{Bonciani:1998vc}, and results
beyond the NLO and NLL level of
accuracy have recently been published
\cite{Moch:2008qy,arXiv:0906.5273,Beneke:2009ye,Ahrens:2010zv,Beneke:2010fm,Kidonakis:2010dk,Ahrens:2011mw,Ahrens:2011px}.

At the current level of precision, the theory agrees
\cite{Moch:2008qy,Ahrens:2010zv,Ahrens:2011mw,Cacciari:2008zb,
  Kidonakis:2008mu} with the data from the Tevatron and
LHC~\cite{Abazov:2011cq,Abazov:2011mi,Aaltonen:2011tm,Aaltonen:2010bs,Khachatryan:2010ez,Collaboration:2011nb,Aad:2010ey,:1900yx,Aad:2011yb,CDF-top-sigma,ATLAS-top-sigma,CMS-top-sigma}. The
large statistics becoming available at the LHC, and the precision of
the experiments, will however soon bring the accuracy of the
measurements to the level of possibly less than 5 percent, thus
challenging the present theoretical systematics.

In this paper we extend to the next-to-next-to-leading logarithmic accuracy
(NNLL) the Mellin $N$-space soft-gluon resummation of the top-pair
production cross-section which was first developed
in~\cite{Catani:1996dj,Catani:1996yz} for leading (LL), and
in~\cite{Bonciani:1998vc} for next-to-leading logarithmic
accuracy (NLL).
This calculation uses the recently derived two-loop
anomalous dimension matrices \cite{Beneke:2009rj,Czakon:2009zw},  and
matches the result to the best known approximation to the NNLO
cross-section~\cite{Beneke:2009ye}. The numbers we derive are
very robust and show a significant stability - as expected - given
that this is the third order at which the soft-gluon resummation is
applied to this observable (i.e. LL, NLL and NNLL).

We include in our assessment the latest sets of parton distribution
function (PDF) fits, analyze the impact of the emerging contributions
at ${\cal O}(\alpha_s^4)$ like two-loop Coulomb corrections and, most
importantly, missing NNLO corrections. Addressing the full theoretical
uncertainty associated with the total $t\bar t$ cross-section as
currently known, we conclude that we see no evidence for a strong
reduction of the theoretical uncertainty compared to the long-ago
established NLO+NLL analysis. Based on our comprehensive analysis we,
however, speculate on the uncertainty that should be achievable once
the full NNLO calculation will be completed, which could be
significant.

The paper is organized as follows: in section \ref{sec:framework} we
summarize the key elements of the threshold approximation at NNLO, and
develop the Mellin-space resummation formalism to the
NNLL level, leaving some technical details to the Appendix. 
In section \ref{sec:syst} we study
the details of the theoretical systematics and the sensitivity of our
theory prediction to the scale dependence and to a number of currently
unknown contributions of ${\cal O}(\alpha_s^4)$ and higher. 
In section \ref{sec:pheno} we include the study of the PDF
systematics, and present our best
predictions for the Tevatron and LHC (7 TeV). 
We present the dependence of the predicted
cross-section on the value of the top mass and compare with the most
precise available experimental measurements,  finding a very
satisfactory agreement. In the concluding section we discuss the
comparison of our results with the current literature.

\section{Theoretical framework}
\label{sec:framework}
\subsection{NNLO results in the threshold approximation}
\label{sec:nnlo-approx}
While the calculation of the full NNLO results for the total cross
section is yet to be completed, recent work has determined its exact
behavior near the production threshold, where it can be represented as
an expansion in powers of $1/\beta$ and $\log\beta$
($\beta=\sqrt{1-4m^2/\hat{s}}$ being the heavy quark velocity in the
$Q\bar{Q}$ rest frame and $m$ the top pole mass). In the limit
$\beta\to 0$ the ${\cal O}(\alpha_s^4)$ partonic cross section can be
written as
\begin{equation}
\sigma^{(2)}_{ij, \mathbf{I}}(\beta,\mu,m) = 
\sigma^{(0)}_{ij, \mathbf{I}}(\beta,\mu,m) \, 
\left( \frac{\alpha_s(\mu^2)}{4\pi} \right)^2
\, \left[
\sigma^{(2,0)}_{ij, \mathbf{I}}(\beta) +
\sigma^{(2,1)}_{ij, \mathbf{I}}(\beta) \, \log\left(\frac{\mu^2}{m^2}\right) +
\sigma^{(2,2)}_{ij, \mathbf{I}}(\beta) \, \log^2\left(\frac{\mu^2}{m^2}\right) 
\right] \; ,
\label{eq:sigma2-color}
\end{equation}
where we defined
$\sigma^{(0)}_{ij, \mathbf{I}}$ as the Born-level cross section.
For ease of notation we set the factorization ($\mu_F$) and
renormalization ($\mu_R$) scales equal to $\mu$, although in our calculation
we allowed them to vary independently.

The index $\mathbf{I} = \mathbf{1}, \mathbf{8}$ corresponds to the color configuration of the
heavy quark pair, $(ij)=(q\bar{q})$, $(gg)$ are the initial-state partons and the functions
$\sigma^{(2,n)}$ are expanded as follows:
\begin{equation}
\label{eq:nnlo}
\sigma^{(2,n)}_{ij,\mathbf{I}}(\beta) = \sum_{a=0}^{2-n}\sum_{b=0 \atop a+b>0}^{4-2a-n}
k^{(n;a,b)}_{ij,\mathbf{I}}~ \frac{1}{\beta^{a}} \,
\log^b{\beta} + C^{(2,n)}_{ij,\mathbf{I}} + {\cal O}(\beta)\; .
\end{equation}
For $n=1,2$ the above functions, including the $\beta$-independent terms
$C^{(2,n)}$, can be derived entirely from the knowledge of the NLO
results. In particular, the $C^{(2,n)}$ ($n=1,2$) terms can be determined by
enforcing the compensation of their scale dependence against
that of appropriate terms of ${\cal O}(\as^3)$. 
For $n=0$, the functions were extracted in~\cite{Beneke:2009ye} from the threshold behavior
of the NNLO result, up to (but excluding) the constant term
$C^{(2,0)}$, which will only become available with the completion of
a full NNLO calculation.

To the extent that the hadronic production of heavy quark pairs
receives an important contribution from the region near threshold, it
is meaningful to incorporate the known singular threshold terms into
an improved prediction of the total production cross section, even
though they do of course not represent the full NNLO result. We label
NNLO$_\beta$~\cite{loopfest2010} the approximation that adds to the full NLO all
${\cal O}(\as^4)$ terms that become singular when $\beta\to 0$.  Note that we
do not impose scale dependence cancellation at ${\cal O}(\as^4)$, i.e. we
exclude all the finite contributions $C^{(2,n)}$: while the
$n=1,2$ terms are known from scale dependence compensation with 
${\cal O}(\as^3)$ terms, the lack
of any threshold enhancement prevents us from assuming that their value
should be bigger than the unknown $C^{(2,0)}$ (This will be confirmed
with direct numerical studies presented in Section~\ref{sec:syst-logON}.). 
Thus we find it more coherent to neglect the non-singular terms.
We also remark that if one wants to include these scale dependent terms,
one should still consider the ambiguity in the choice of the scale that
divides the renormalization and factorization scale in the arguments of the
logarithms. These scales are uniquely fixed only if the constant terms
are known. Thus, the compensation of the renormalization and factorization
scale variations induced by these terms will be counterbalanced by the
uncertainties obtained by varying these new scales. It is therefore simpler
not to include these terms at all, and let the lack of scale compensation
work as an indicator of unknown higher order terms. Our approach is
therefore to account for the ignorance of the ${\cal O}(\as^4)$
constants through the scale variation systematics. In addition, we
shall show explicitly in the following that two different choices of
these constants,  $C^{(2,0)}=0$ and  $C^{(2,0)}=\overline{C}^{(2,0)}$
(see Appendix), lead to differences consistent with the scale systematics.

\subsection{Soft-gluon resummation with NNLL accuracy}\label{sec:N-space}
We extend here the NLO+NLL formalism for soft gluon resummation of the
total top-pair hadroproduction cross-section, discussed in
Ref.~\cite{Bonciani:1998vc}, to include the resummation of NNLL terms.

For simplicity, we first summarize the qualitative overall structure of
our result:
\begin{itemize}
\item the ${\cal O}(\as^3)$ contributions correspond to
the exact NLO result;
\item We  perform the resummation in Mellin space and then invert numerically
back to $x$-space using the Minimal Prescription of
Ref.~\cite{Catani:1996yz}.
\item the truncation of the NNLL result at ${\cal O}(\as^4)$ includes
  all singular contributions described by NNLO$_\beta$, plus non
  singular terms that arise from the inverse Mellin transform of the
  $N$-space resummation; in particular, terms of ${\cal O}(1/N)$ can
  give non-negligible contributions, which reflect the uncertainty
  about higher-order non-singular terms.
This point in particular underscores a qualitative and quantitative difference with the
alternative approach of simply matching at ${\cal O}(\as^4)$ the
resummed result to NNLO$_\beta$.

\end{itemize}
The resummed cross-section in $N$-space\footnote{The Mellin moments $N$
of a function $g(\rho)$, with $\rho = 4m^2/\hat{s}$, are defined by
$g_N=\int_0^1 d\rho \, \rho^{N-1} \, g(\rho)$.} reads:
\beqn 
\sigma_N^{(res)}(m^2) &=& \sum_{ij=q{\bar q},gg} \;F_{i,N+1}(\mu^2) \;F_{j,N+1}(\mu^2)
\left[ {\hat \sigma}_{ij,N}^{(res)}(m^2, \as(\mu^2),\mu^2) - \left( {\hat \sigma}_{ij,N}^{(res)}(m^2, \as(\mu^2),\mu^2)
\right)_{\as^3} \right] \nn \\
\label{HQCSNres}
&+& \sigma_N^{({\rm NLO})}(m^2) \;\;,
\eeqn
where $\sigma_N^{({\rm NLO})}$ is the NLO hadronic cross-section
\cite{Nason:1987xz,Beenakker:1988bq,Czakon:2008ii}, ${\hat
  \sigma}_{ij,N}^{(res)}$ is the NNLL resummed partonic cross section,
and $\left( {\hat \sigma}_{ij,N}^{(res)} \right)_{\as^3}$ is
its perturbative truncation at order $\as^3$. As before, we set
here the renormalization ($\mu_R$) and factorization ($\mu_F$) scales
equal to $\mu$: they are however kept separate and varied
independently in the subsequent studies of the scale systematics.

In the threshold limit $N\to\infty$ the NNLL resummed partonic cross section ${\hat\sigma}_{ij,N}^{(res)}$ factorizes:
\beq                    
\label{fijress} 
{\hat \sigma}_{ij, \, N}^{(res)}(m^2,\as(\mu^2),\mu^2) = \sum_{{\bf I = 1,8}}
{\hat \sigma}_{ij,{\bf I}, \, N}^{(Coul)}(\as(\mu^2),\mu^2/m^2) \; 
{\hat \sigma}_{ij,{\bf I}}^{(Hard)}(\as(\mu^2),\mu^2/m^2) \; 
\Delta_{ij,{\bf I}, \,N+1}\!\left( \as(\mu^2), \frac{\mu^2}{m^2} \right) \, ,
\eeq
in terms of the radiative factors $\Delta_{ij,{\bf I}, \, N}$
containing all contributions due to soft-gluon emission, the functions
${\hat \sigma}_{ij,{\bf I}, \, N}^{(Coul)}$ containing the
threshold-enhanced bound-state contributions (Coulomb effects) and the
hard matching functions ${\hat \sigma}_{ij,{\bf I}}^{(Hard)}$. 

The radiative factors introduced in Eq.~(\ref{fijress}),
$\Delta_{ij,{\bf I}, \, N}$, are given by:
\ba
\ln \Delta_{ij,{\bf I}, \,N} &=& \int_{0}^{1} dz \, \frac{z^{N-1}-1}{1-z}
\,  \left\{ \int_{\mu^2}^{4m^2(1-z)^2}
\frac{dq_{\perp}^2}{q_{\perp}^2}
\, 2\delta_{ij} A_{i} \left( \as(q_{\perp}^2) \right) 
+ D_{ij,{\bf I}} \left( \as \left( 4m^2(1-z)^2 \right) \right) \right\} 
\nn \\
&+& {\cal O}(\as (\as \ln N)^k) \; . \label{dqq8}
\ea

The process independent anomalous dimensions $A_{i}$ describe
soft-collinear initial state radiation. They are known through three
loops \cite{KT,CdET,Moch:2004pa}; the explicit expressions can be
found in Ref.~\cite{Moch:2004pa}. The anomalous dimensions $D_{ij,{\bf
    I}}$ describe wide-angle soft radiation and depend both on the
initial and final states~\cite{bonciani}. For top-pair production, 
they have recently been
derived through two loops in
Refs.~\cite{Beneke:2009rj,Czakon:2009zw}. Their explicit expressions
read:
\ba
D_{q{\bar q},{\bf 8}} &=& -C_A {\alpha_S(\mu^2)\over \pi} + \left({\alpha_S(\mu^2)\over \pi}\right)^2\Bigg\{ 
\left(-{115\over 36}+{\pi^2\over 12}-{\zeta_3\over 2}\right)C_A^2 + \left( -{101\over 27} + {11\pi^2\over 18} + {7\zeta_3\over 2}\right)C_AC_F\nonumber\\
 && +{11\over 18}C_A N_F + \left( {14\over 27}- {\pi^2\over 9}\right)C_FN_F\Bigg\} +{\cal O}(\alpha_S^3)\, , \nonumber\\ 
 D_{gg,{\bf 8}} &=& -C_A {\alpha_S(\mu^2)\over \pi} + \left({\alpha_S(\mu^2)\over \pi}\right)^2\Bigg\{ 
\left(-{749\over 108}+{25\pi^2\over 36}+3\zeta_3\right)C_A^2 + \left( {61\over 54}- {\pi^2\over 9}\right)C_AN_F\Bigg\} +{\cal O}(\alpha_S^3)\, , \nonumber\\
 D_{gg,{\bf 1}} &=& \left({\alpha_S(\mu^2)\over \pi}\right)^2\Bigg\{ 
\left(-{101\over 27}+{11\pi^2\over 18}+{7\zeta_3\over 2}\right)C_A^2 + \left( {14\over 27} - {\pi^2\over 9}\right)C_AN_F\Bigg\} +{\cal O}(\alpha_S^3) \, ,
\label{eq:anomdim}
\ea
where $N_F$ is the number of light flavours, i.e. lighter than the top quark.
The integrations in Eq.~(\ref{dqq8}) can be performed analytically and the resummed NNLL cross-section can be written explicitly. To extend the NLL results of Ref.~\cite{Bonciani:1998vc}, one can utilize, for example, the results in Ref.~\cite{Vogt:2000ci}.

The $N$-independent, hard matching functions read: 
\begin{equation}
{\hat \sigma}_{ij,{\bf I}}^{(Hard)}(\as(\mu^2),\mu^2/m^2)  = \left(\as(\mu^2)\right)^2\left( 1 + {\as(\mu^2)\over \pi} H_{ij,{\bf I}}^{(1)}(\mu^2/m^2)  + \left({\as(\mu^2)\over \pi}\right)^2 H_{ij,{\bf I}}^{(2)}(\mu^2/m^2) + {\cal O}(\as^3) \right) \, .
\label{eq:Hard}
\end{equation}
The one-loop results $H_{ij,{\bf I}}^{(1)}(\mu^2/m^2)$ have been
calculated in Ref.~\cite{Czakon:2008cx}. The
renormalization/factorization scale dependence of the two-loop
corrections $H_{ij,{\bf I}}^{(2)}(\mu^2/m^2)$ can be obtained, for
example, from the results of Ref.~\cite{Beneke:2009ye}:
\begin{eqnarray}
H_{q{\bar q},{\bf 8}}^{(2)}(\mu^2/m^2)  &=& H_{q{\bar q},{\bf 8}}^{(2)}(1) + 8.91918 \ln^2\left({\mu^2\over m^2}\right) + 34.7212 \ln\left({\mu^2\over m^2}\right) \, , \nonumber\\
H_{gg,{\bf 8}}^{(2)}(\mu^2/m^2)  &=& H_{gg,{\bf 8}}^{(2)}(1) + 9.31619 \ln^2\left({\mu^2\over m^2}\right) + 60.2080 \ln\left({\mu^2\over m^2}\right) \, ,\nonumber\\
H_{gg,{\bf 1}}^{(2)}(\mu^2/m^2)  &=& H_{gg,{\bf 1}}^{(2)}(1) + 9.31619 \ln^2\left({\mu^2\over m^2}\right) + 38.5239 \ln\left({\mu^2\over m^2}\right) \, .
\label{eq:H2}
\end{eqnarray}
The genuine two-loop constants $H_{ij,{\bf I}}^{(2)}(1)$ are
currently unknown, and are related to the constants
$C^{(2,0)}_{ij,\mathbf{I}}$ introduced in (\ref{eq:nnlo}), as
discussed in the Appendix. For simplicity, in Eq.~(\ref{eq:H2}) we have given
directly the numerical values of the scale dependent term evaluated
for $N_F=5$. In analogy with our discussion of the  
finite $C^{(2,n)}$ coefficients, which we suppress in the
NNLO$_\beta$ approximation, we shall set $H_{ij,{\bf
    I}}^{(2)}(\mu^2/m^2)=0$ in our default NLO+NNLL predictions, as
will be motivated by the numerical results of
Section~\ref{sec:syst-logON}.

To analyze the impact of subleading, power-suppressed corrections we
follow Ref.~\cite{Bonciani:1998vc} and introduce a power suppressed
term controlled by a constant $A$ into the function $H_{ij,{\bf
    I}}^{(1)}$. The choice $A=0$ sets this additional term to zero,
while the default value $A=2$ was chosen in
Ref.~\cite{Bonciani:1998vc}. The rationale behind the inclusion of
this term was the observation that power suppressed terms ${\cal
  O}(1/N)$ are needed to bring the resummed cross-section closer to
the fixed order NLO result away from the threshold region. We will
have more to say about the numerical impact of this term in
Section~\ref{sec:syst}.

In deriving the Coulomb contributions ${\hat \sigma}_{ij,{\bf I}, \,
  N}^{(Coul)}$ we follow the approach of Ref.~\cite{Bonciani:1998vc}
and absorb in it the exact Born cross-section, except for its overall
factor of $\as^2$ that we attribute to the hard function - see
Eq.~(\ref{eq:Hard}). The functions ${\hat
  \sigma}_{ij,{\bf I}, \, N}^{(Coul)}(\as(\mu_R^2),\mu_R^2/m^2)$,
which only depend on the renormalization scale, are
obtained as the Mellin transform of the following $x$-space functions:
\ba
{\hat \sigma}_{ij,{\bf I}}^{(Coul)}(\rho, \as(\mu_R^2),\mu_R^2/m^2) &=& {{\hat \sigma}_{ij,{\bf I}}^{(Born)}(\rho, \as(\mu_R^2))\over \as^2(\mu_R^2)}\nonumber\\
&\times&\Bigg\{1+{\alpha_S(\mu_R^2)\over \pi}C^{(1)}_{ij,{\bf I}}(\rho) + \left({\alpha_S(\mu_R^2)\over \pi}\right)^2C^{(2)}_{ij,{\bf I}}(\rho,\mu_R^2/m^2) +{\cal O}(\alpha_S^3)\Bigg\} \, .
\label{eq:Coulomb}
\ea

The Born cross-sections, for all reactions and color configurations,
have been given in Ref.~\cite{Bonciani:1998vc}, together with the
one-loop Coulomb functions $C^{(1)}_{ij,{\bf 1}}(\rho) =
C_F\pi^2/(2\beta)$ and $C^{(1)}_{ij,{\bf 8}}(\rho) =
(C_F-C_A/2)\pi^2/(2\beta)$. The two-loop functions can be extracted
from the results of Ref.~\cite{Beneke:2009ye} by matching them to the
Mellin-inverse of Eq.~(\ref{fijress}):
\ba
C^{(2)}_{gg,{\bf 1}}(\rho,\mu_R^2/m^2) &=& {C_F^2 \pi^4\over 12 \beta^2} -C_F\left(C_F+{C_A\over 2}\right) 2\pi^2 \ln(\beta) \nonumber\\
&& + {\pi^2\over \beta}\Bigg\{  \left( -{11\over 12}C_FC_A + {C_F N_F\over 6}\right) \ln\left({2m\,\beta \over \mu_R}\right)  +{31\over 72} C_FC_A -{5\over 36} C_F
N_F \Bigg\}\, ,  \nonumber\\  
C^{(2)}_{gg,{\bf 8}}(\rho,\mu_R^2/m^2) &=&C^{(2)}_{gg,{\bf 1}}(\rho,\mu_R^2/m^2) \left[C_F\rightarrow C_F-{C_A\over 2}\right] \, , \nonumber\\
C^{(2)}_{q{\bar q},{\bf 8}}(\rho,\mu_R^2/m^2) &=&C^{(2)}_{gg,{\bf 8}}(\rho,\mu_R^2/m^2)  + \left(C_F-{C_A\over 2}\right)^2 {4\pi^2\over 3} \ln(\beta)  \, .
\label{eq:2-loop-Coulomb}
\ea

The Mellin transform of Eq.~(\ref{eq:Coulomb}) through NLO has been
calculated in Ref.~\cite{Bonciani:1998vc}. We do not present here the
Mellin transform of the NNLO corrections. They are given by large
expressions that are straightforward to calculate following the
discussion of Ref.~\cite{Bonciani:1998vc} and utilizing the following
approximation:
\be
\ln\left( {1+\beta\over 1-\beta}\right) \approx  - \ln(\rho) + 2\left( 0.9991\beta -0.4828\beta^2 + 0.2477\beta^3 -0.0712\beta^4\right) \, .
\ee

Before concluding this section 
we stress again that in Eq.~(\ref{HQCSNres}) we use the NLO fixed order
result $\sigma^{({\rm NLO})}(m^2)$ and not $\sigma^{({\rm
    NNLO}_\beta)}(m^2)$ as one might expect. The reason is that
all the information to be found in the approximate NNLO cross-section
$\sigma^{({\rm NNLO}_\beta)}$ is already contained, {\it by
  matching}, in the all order resummed result ${\hat
  \sigma}_{ij,N}^{(res)}$. In particular, the two loop anomalous
dimensions in Eq.~(\ref{eq:anomdim}) control the single $\ln\beta$
terms in $\sigma^{({\rm NNLO}_\beta)}$, and the two-loop Coulomb
terms in Eq.~(\ref{eq:Coulomb}), including the potentials
$\sim\ln\beta$, have been matched to the Coulomb terms in
$\sigma^{({\rm NNLO}_\beta)}$.

To make this point completely transparent we note that the difference between
Eq.~(\ref{HQCSNres}) and its analogue defined by using
$\sigma^{({\rm NNLO}_\beta)}$ instead (and, of course, subtracting
the terms in ${\hat \sigma}_{ij,N}^{(res)}$ through ${\cal O}(\as^4$))
is given in N-space by the following expression:
\begin{equation}
\sum_{ij=q{\bar q},gg} \;F_{i,N+1} \;F_{j,N+1} \left[ {\hat \sigma}_{ij,N}^{(res)}\Big\vert_{{\cal O}(\as^4)~{\rm only}}\right] ~ - ~
\sigma_N^{{\rm NNLO}_\beta}\Big\vert_{{\cal O}(\as^4)~{\rm only}} \, .
\label{eq:differenseas4}
\end{equation}
In other words, Eq.~(\ref{eq:differenseas4}) represents the difference
between the terms of order $\as^4$ derived respectively within the
resummed $N$-space and the fixed order $x$-space approaches. Since, as
we just explained, the two contain the same input they do cancel each
other, at least in the limit $N\to \infty$. In practice
Eq.~(\ref{eq:differenseas4}) contains power-suppressed terms that
behave as ${\cal O}(1/N)$ in the soft limit $N\to\infty$. These power
suppressed terms originate in the lower loop (LO and NLO) terms in
Eq.~(\ref{fijress} ) and, as it turns out, are not numerically
negligible. Given that both terms in Eq.~(\ref{eq:differenseas4}) are
the result of an approximation, and as we favour $N$-space resummation
since it's less likely to introduce large terms due to the violation
of momentum conservation, we perfer not to introduce the
power terms Eq.~(\ref{eq:differenseas4}) into
Eq.~(\ref{HQCSNres}). Such ${\cal O}(1/N)$ ambiguity is inherent in
the soft approximation irrespective of the details of its
implementation and can only be removed by adding to
Eq.~(\ref{HQCSNres}) the full NNLO result, once it becomes available.

\section{Study of the theoretical systematics}
\label{sec:syst}
In this section we focus on the purely theoretical systematics,
arising from the scale dependence of the cross sections, and from the
different possible descriptions of higher-order terms not controlled
by resummation, like unknown two-loop (threshold) hard matching
constants $H_{ij,{\bf I}}^{(2)}(1)$ and terms vanishing at
threshold. We shall then complete the study of systematics, including PDF
and mass dependence, in Section~\ref{sec:pheno}.

\subsection{Benchmark results}
We shall compute reference values for our $t\bar t$ cross section
predictions at $m_t = 173.3$~GeV~\cite{:1900yx}\footnote{The best
  measured value for the top mass has been recently updated in
  Ref.~\cite{Lancaster:2011wr} to $173.2\pm0.9$~GeV (for a recent
  review see~\cite{arXiv:1109.2163}). However, for use
  here the previously published value of 173.3 to facilitate the
  comparison with other recent theoretical analyses that used this
  $m_t$, such as~\cite{Beneke:2011mq}. We estimate (see
  section~\ref{sec:pheno}) that the change of 0.1 GeV, from 173.3 to
  173.2 GeV, would lead to an increase of about 0.3\% in the cross section
  values, well within the overall theoretical uncertainties. \label{footnote}}.
The central values of these predictions are obtained for $\mu_R=\mu_F=m_t$.  Throughout the paper we use the
strong coupling constant evaluated at scale $\mu_R$ as provided by the
corresponding PDF set.  Our default parton
distribution set for NLO (and NLO+NLL) is MSTW2008nlo68cl, whereas for
NNLO$_\beta$ and NNLL resummed calculations we use the
MSTW2008nnlo68cl set~\cite{Martin:2009iq}. In all cases we include
them through the LHAPDF interface~\cite{Whalley:2005nh}.

The scale systematics is evaluated by varying the renormalization and
factorization scales independently in the range suggested in
ref.~\cite{Cacciari:2008zb}:
\be \label{eq:musyst}
m_t/2 < \mu_R, \mu_F < 2 m_t, \quad \mathrm{with} \quad 1/2 < \mu_R/\mu_F < 2 \; ,
\ee
and searching for the minimum/maximum of the resulting cross-section. 
It is usually sufficient to consider only the
endpoints of the range Eq.~(\ref{eq:musyst}), namely the pairs 
$(\mu_r/m_t,\mu_F/m_t)=(2,1)$, $(0.5,1)$, $(1,2)$, $(1,0.5)$,
$(2,2)$ and $(0.5,0.5)$ . We have verified that
a search over a grid
with a few hundred points satisfying Eq.~(\ref{eq:musyst}) in the
$(\mu_F , \mu_R)$ plane agrees to within few per mille with the minimum
and maximum rates found in the scan of the endpoints.

Different power-suppressed terms are probed by varying the parameter
$A$ over the two values $A=0$ and $A=2$, as discussed
in~\cite{Bonciani:1998vc}.  

%
\begin{table}[ht]
\begin{center}
\begin{tabular}{| c | l | c | c | c | c |}
\hline
& {\rm Approximation}  & $\sigma_{\rm tot}~[{\rm pb}]$ &  PDF & A  &
pure 2-loop Coulomb \\
\hline 
1 & NLO     & $6.681^{+0.363\, (5.4\%)}_{-0.752\, (11.3\%)}$ & NLO & --  & --  \\ 
\hline
2 & NLO+NLL & $7.070^{+0.212\, (3.0\%)}_{-0.432\,(6.1\%)}$ & NLO & 0 & --  \\ 
\hline
3 & NLO+NLL & $6.930^{+0.278\, (4.0\%)}_{-0.496\,(7.2\%)}$ & NLO & 2 & --  \\ 
\hline
4 & NNLO$_\beta$,  ${C}^{(2,0)}_{ij}=0$  & $7.062^{+0.240\, (3.4\%)}_{-0.334\, (4.7\%)}$  & NNLO & -- & --  \\ 
\hline
5 & NNLO$_\beta$, ${C}^{(2,0)}_{ij}=\overline{C}^{(2,0)}_{ij}$  & $6.853^{+0.268\, (3.9\%)}_{-0.386\, (5.6\%)}$  & NNLO & -- & --\\ 
\hline
6 & NLO+NNLL & $6.844^{+0.197\, (2.9\%)}_{-0.353\,(5.2\%)}$ & NNLO & 0
& NO \\
\hline
7 & NLO+NNLL & $6.722^{+0.212\, (3.2\%)}_{-0.391\,(5.8\%)}$ & NNLO & 2
& NO \\
\hline
8 & NLO+NNLL & $6.844^{+0.215\, (3.1\%)}_{-0.377\,(5.5\%)}$ & NNLO & 0
& YES \\
\hline
\color{blue}{9} & \color{blue}{NLO+NNLL} & \color{blue}{ $6.722^{+0.243\, (3.6\%)}_{-0.410\,(6.1\%)}$ }& \color{blue}{NNLO} & \color{blue}{2} & \color{blue}{YES}\\
\hline

\end{tabular}
\caption{\label{tab:tev-syst} Central values and theoretical systematics for the various
  approximations to $\sigma_{\rm tot}$, in pb, at the
  Tevatron. $m_{\rm top}=173.3$ GeV, PDF$_{\rm NLO}$=MSTW2008nlo68cl,
  PDF$_{\rm NNLO}$=MSTW2008nnlo68cl. Row 9, highlighted in blue, gives
  our best prediction for central value and scale systematics. The predicted
  cross-sections are presented, if applicable, depending on the values
  of the constant $A$ and on whether {\it pure}
two-loop Coulomb corrections
  (\ref{eq:2-loop-Coulomb}) are included or not.}
\end{center}
\end{table}

\begin{table}[ht]
\begin{center}
\begin{tabular}{| c | l | c | c | c | c |}
\hline
& {\rm Approximation}  & $\sigma_{\rm tot}~[{\rm pb}]$ &  PDF & A  &
pure 2-loop Coulomb \\
\hline
1 & NLO    & $158.1^{+19.5\, (12.3\%)}_{-21.2\, (13.4\%)}$ & NLO & -- & --\\ 
\hline
2 & NLO+NLL & $174.8^{+17.6\, (10.1\%)}_{-15.3\, (8.8\%)}$ & NLO & 0 & -- \\
\hline
3 & NLO+NLL & $167.1^{+14.3\, (8.6\%)}_{-15.4\, (9.2\%)}$ & NLO & 2 & -- \\
\hline
4 & NNLO$_\beta$, ${C}^{(2,0)}_{ij}=0$  & $ 161.2^{+11.3\, (7.0\%)}_{-10.8\, (6.7\%)}$  & NNLO & -- & -- \\
\hline
5 & NNLO$_\beta$, ${C}^{(2,0)}_{ij}=\overline{C}^{(2,0)}_{ij}$ 
 & $ 154.0^{+12.0\, (7.8\%)}_{-8.6\, (5.6\%)}$  & NNLO & --  & -- \\
\hline
6 & NLO+NNLL & $161.5^{+14.5\, (9.0\%)}_{-12.3\,(7.6\%)}$ & NNLO & 0 & NO \\
\hline
7 & NLO+NNLL & $155.9^{+11.5\, (7.4\%)}_{-13.0\,(8.3\%)}$ &  NNLO & 2 & NO \\
\hline
8 & NLO+NNLL & $164.7^{+15.0\, (9.1\%)}_{-12.8\,(7.8\%)}$ &  NNLO & 0 &YES\\
\hline
\color{blue}{9} & \color{blue}{NLO+NNLL} & \color{blue}{
  $158.7^{+12.2\, (7.7\%)}_{-13.5\,(8.5\%)}$ }&\color{blue}{NNLO}&\color{blue}{ 2 }&\color{blue}{ YES } \\
\hline
\hline
\end{tabular}
\caption{\label{tab:lhc7-syst} As in Table~\ref{tab:tev-syst} but for the LHC at 7 TeV.}
\end{center}
\end{table}
%
Our numerical results are summarized in tables \ref{tab:tev-syst} and \ref{tab:lhc7-syst}. 
As a benchmark, in rows 1-3 of these tables we present the well-understood NLO and
NLO+NLL ($A=0$ and $A=2$) results. They are an update of
Ref.~\cite{Bonciani:1998vc}.

In row 4 we present the NNLO$_\beta$ approximation as defined in
Section ~2,
and in row 5 we show the effect of including non-zero values for the
constants ${C}^{(2,0)}_{ij}$, setting them to the value 
$\overline{C}^{(2,0)}_{ij}$ defined in the Appendix
in Eq.~(\ref{eq:Cbar}). We notice that, while the numerical impact of
including these constants is noticeable, it is smaller than the
overall scale uncertainty. Therefore we argue that the scale variation
can largely account for the uncertainty stemming from unknown part of the higher-order terms.

At order $\as^4$ the fixed order NNLO approximation contains terms that
are not predicted by the NLO+NLL result, like NNLL soft-enhanced terms and
{\it pure} two-loop Coulomb terms, i.e. $\as^4$ Coulomb terms that do not
arise in the expansion of Eq.~(\ref{fijress}) from the product of
one-loop contributions to ${\hat 
  \sigma}_{N}^{(Coul)}$ with ${\hat \sigma}^{(Hard)}\Delta_{N+1}$. 
It lacks however terms at ${\cal O}(\as^5)$
and beyond that are contained in the resummed results. 
The NLO+NNLL approximation combines both these ingredients, and is
 superior to the NNLO$_\beta$ one, since it
contains all the information to be found in NNLO$_\beta$, plus the
towers of soft LL, NLL and NNLL logs beyond order ${\cal
  O}(\as^4)$. 
The NLO+NNLL rates are given in
 rows 6-9, where we also describe the impact
of subleading $1/N$ terms (through the constant $A$), and of the two-loop
Coulomb effects that were absent in the NLO+NLL results of
Ref.~\cite{Bonciani:1998vc}.

We observe that these two-loop Coulomb terms
(\ref{eq:2-loop-Coulomb}) have a sub-per-mille effect
on the central values for the Tevatron. At the LHC the effect is larger,
of order 2$\%$.  Their effect on
the scale uncertainty is at most at the few per-mille level and is
thus negligible (for both Tevatron and LHC). We do not resum the
Coulombic corrections beyond order $\as^4$. This resummation has been
performed in Ref.~\cite{Beneke:2011mq} and the effect was found to be
negligible.

From tables \ref{tab:tev-syst} and \ref{tab:lhc7-syst} we also observe
a dependence of the predicted cross-section on the value of
the constant $A$. As we emphasized before, we consider the inclusion
of non-zero $A$ as a model for the power suppressed terms $\sim 1/N$
that are not controlled by the threshold approximation. We observe a
modest $0.5\%$ decrease in scale dependence from including $A\neq 0$
and a $2\%$ shift of the central
value, which is consistent with the overall scale systematics 
(the corresponding changes for the LHC are $0.5\%$ and $3\%$
respectively). 
We conclude that, while our estimate of the size of the power
suppressed terms is not comprehensive, it clearly shows that power
suppressed terms can be a significant, few-percent-effect on the
central value both at the Tevatron and LHC.

\begin{figure}[t]
  \begin{center}
   \includegraphics[width=8.3cm]{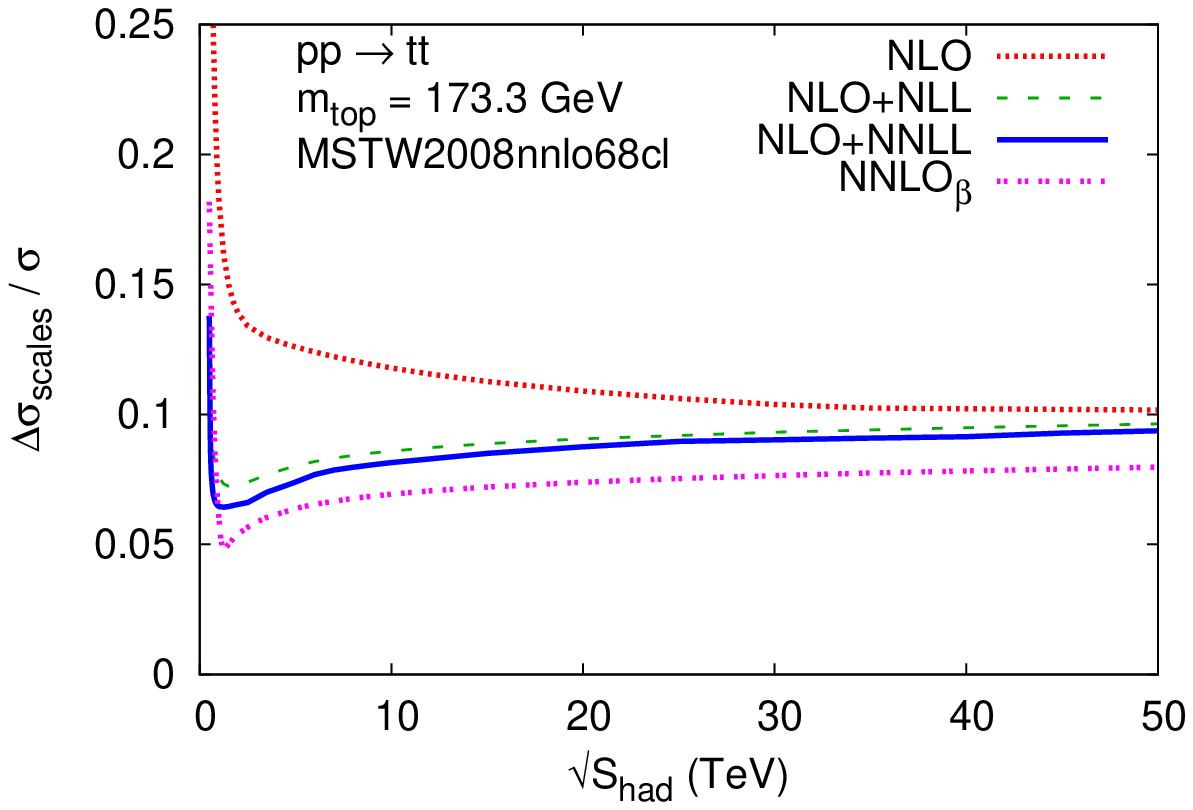}
   \hspace{-.4cm} 
   \includegraphics[width=8.3cm]{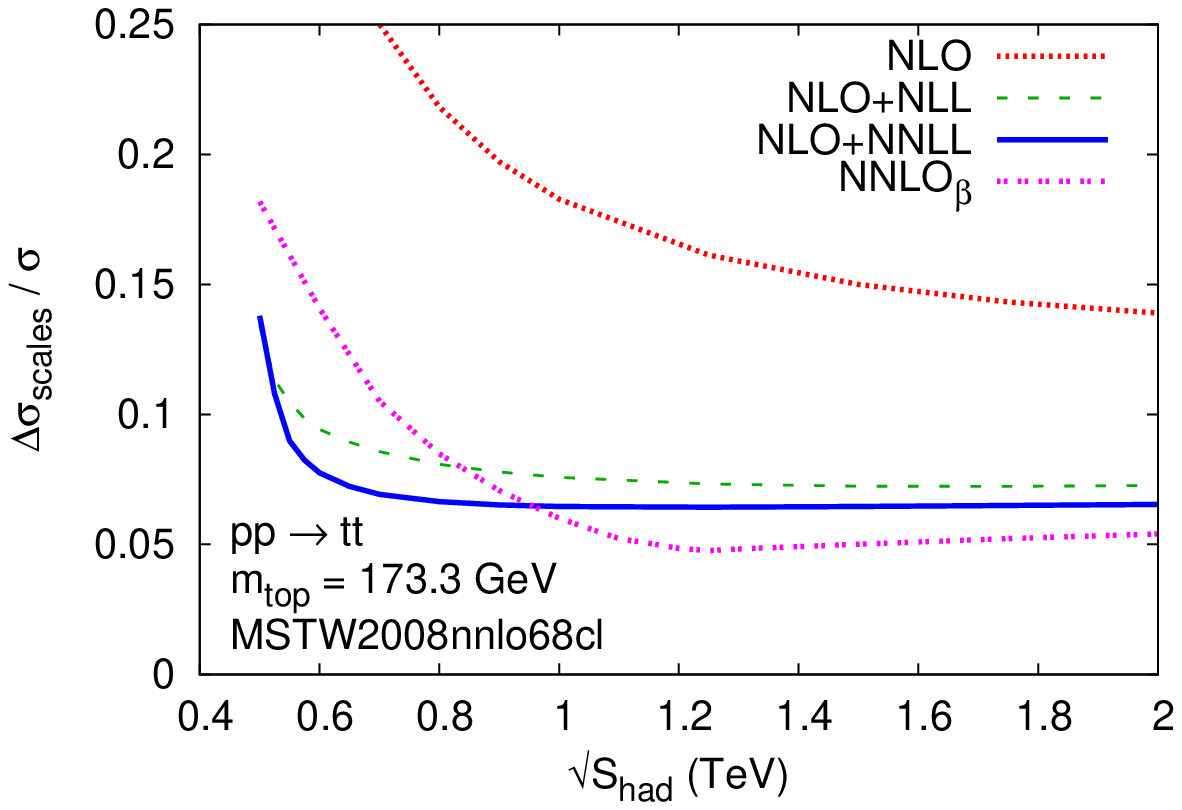} 
   \caption{  \label{fig:uncertainty-S-had} Relative half uncertainty in the total inclusive $t\bar t$ 
   cross-section 
   as a function of the centre-of-mass energy $\sqrt{S_\mathrm{had}}$ of a proton-proton collider.
   Left plot: whole energy range up to $\sqrt{S_\mathrm{had}}=50$~TeV. 
   Right plot: blow-up of the threshold region, up to $\sqrt{S_\mathrm{had}}=2$~TeV.
     }
  \end{center}
\end{figure}

Before closing this section we offer an alternative graphical
representation of the
scale uncertainty of the various scenarios given in tables \ref{tab:tev-syst} and
\ref{tab:lhc7-syst}. In Fig.~\ref{fig:uncertainty-S-had} we plot the
relative half uncertainty due to scale variations (defined as 
$\Delta\sigma_\mathrm{scales}/\sigma \equiv (\sigma_{\rm
  max}-\sigma_{\rm min})/(2\sigma_{\rm central})$, so that one can quote for
  the cross section $\sigma_{\rm central} \pm \Delta \sigma_\mathrm{scales}$) 
  as a function of the
hadronic centre of mass energy for $pp$ collisions and for a top mass of
173.3 GeV. As expected, the uncertainty of the NLO calculation is
largest when the production is closest to threshold, and decreases for
larger $\sqrt{S_{\rm had}}$. Upon resummation of the threshold
logarithms, the biggest improvement (i.e. reduction in uncertainty)
can likewise be obtained close to the threshold. One can easily observe the 
better stability of the NLO+NLL and NLO+NNLL results when the threshold is 
approached.
As the $t\bar t$
production takes place at larger $\sqrt{S_{\rm had}}$ the effect of
resummation is reduced, as expected, since the resummed logarithms
become smaller. When $S_{\rm had} \gg
4m^2$ one does not expect any significant improvement from a
resummation of logarithms that are strongly suppressed. Indeed, one
observes from the plot that the uncertainty of the resummed results is
practically identical to that of the fixed order calculation in this
limit\footnote{The NNLO$_\beta$ result seems to display a slightly smaller
uncertainty than the resummed ones. The difference is however likely 
not significant and, in particular, it should not be considered as
suggestive that this approximation constitutes a better prediction.}.

\subsection{Impact of scale-dependent finite terms at \texorpdfstring{${\cal O}(\as^4)$}{oas4}}
\label{sec:syst-logON}
We discuss here the effect of the ${\cal
  O}(\as^4)$  constant terms ${C}^{(2,n)}_{ij}$,
which we introduced in (\ref{eq:nnlo}) and which we suggested should
not be incorporated in either NNLO$_\beta$ or in the NNLL resummed
results. Their main impact is the large reduction in the scale
variation of the cross section,
due to the scale logarithms present for $n=1,2$, whose coefficients
 compensate by construction the scale dependence of ${\cal
  O}(\as^3)$  terms. 
If we focus on the NNLO$_\beta$ case, the addition of
the known ${C}^{(2,n)}_{ij}$ contributions ($n=1,2$), for different
choices of the unknown $n=0$ terms, leads to the following
results:
\ba
\label{eq:logON_1}
\mathrm{Tevatron}: &\quad& \sigma^{\mathrm{NNLO}_\beta^{C^{(2)}}}=6.853^{+0.092\, (1.3\%)}_{-0.408\, (6.0\%)}  ~\mathrm{pb}
\quad \mathrm{for} \quad{C}^{(2,0)}_{ij}=\overline{C}^{(2,0)}_{ij}
\\
\label{eq:logON_2}
                   &\quad& \sigma^{\mathrm{NNLO}_\beta^{C^{(2)}}}=7.062^{+0.064\, (0.9\%)}_{-0.262\, (3.7\%)}  ~\mathrm{pb}
\quad \mathrm{for} \quad{C}^{(2,0)}_{ij}=0
\\
\label{eq:logON_3}
\mathrm{LHC}: &\quad& \sigma^{\mathrm{NNLO}_\beta^{C^{(2)}}}=154.0^{+2.8\, (1.8\%)}_{-3.7\, (2.4\%)} ~\mathrm{pb}
\quad \mathrm{for} \quad{C}^{(2,0)}_{ij}=\overline{C}^{(2,0)}_{ij}
\\
\label{eq:logON_4}
              &\quad& \sigma^{\mathrm{NNLO}_\beta^{C^{(2)}}}=161.2^{+2.1\, (1.3\%)}_{-4.7\, (2.9\%)} ~\mathrm{pb}
\quad \mathrm{for} \quad{C}^{(2,0)}_{ij}=0
\ea
The central values coincide with those obtained in absence of the
${C}^{(2,n)}_{ij}$ ($n=1,2$) terms, since the logarithms vanish at the
central value $\mu=m$. However, the scale dependence is much smaller
than in rows 4 and 5 of our previous tables. At the Tevatron (LHC),
the scale dependence drops from about $\pm 5\%$ ($\pm 7\%$) to about
$\pm 3\%$ ($\pm 2\%$).  This significant reduction clashes however with
the comparable or larger cross section variations ($3\%$ at the
Tevatron and $4\%$ at the LHC) induced by the variation of
${C}^{(2,0)}_{ij}$ within an a priori reasonable range.  We conclude
that the significant reduction in scale dependence in presence of
${C}^{(2,0)}_{ij}$ ($n=1,2$) terms cannot be interpreted as a genuine
reduction in the total theoretical uncertainty, unless it is combined
with the systematics emerging from the unknown value of the two-loop
constants ${C}^{(2,0)}_{ij}$.

Similar conclusions can be drawn from the study of the NLO+NNLL
results. From the viewpoint of threshold expansion, in the limit $N\to
\infty$ one has to neglect the unknown constants $H_{ij,{\bf
    I}}^{(2)}(1)$ and the  $N$-independent,
$\mu_F,\mu_R$-dependent logarithmic terms, i.e. the whole two-loop
hard function $H_{ij,{\bf I}}^{(2)}(\mu^2/m^2)$. This is completely
analogous to what happens in the NNLO$_\beta$ approximation. In
case the scale dependent terms in $H_{ij,{\bf I}}^{(2)}(\mu^2/m^2)$
are retained, the theoretical uncertainty should include the 
variation of the unknown two-loop constant $H_{ij,{\bf
    I}}^{(2)}(1)$, again as in the NNLO$_\beta$ approximation.
For the sake of documentation, we quote here the relevant
results for the  scale
dependence  obtained after inclusion of the hard function $H_{ij,{\bf
    I}}^{(2)}(\mu^2/m^2)$, exploring as an example the two cases of
${H}_{ij,{\bf I}}^{(2)}(1) =0 $ and
${H}_{ij,{\bf I}}^{(2)}(1) =\overline{H}_{ij,{\bf I}}^{(2)}(1) $
introduced in the Appendix:
\ba
\mathrm{Tevatron}: 
&\quad& \sigma^{\mathrm{NLO}+\mathrm{NNLL}^{H^{(2)}}}=6.722^{+0.017\,
  (0.3\%)}_{-0.320\,(4.8\%)} ~\mathrm{pb} \quad \mathrm{for} \quad H_{ij,{\bf
    I}}^{(2)}(1) = 0 
\\
&\quad& \sigma^{\mathrm{NLO}+\mathrm{NNLL}^{H^{(2)}}}=6.968^{+0.009\,
  (0.1\%)}_{-0.224\,(3.2\%)} ~\mathrm{pb} \quad \mathrm{for} \quad  H_{ij,{\bf
    I}}^{(2)}(1) = \overline{H}_{ij,{\bf I}}^{(2)}(1) 
\\
\mathrm{LHC}: 
&\quad& \sigma^{\mathrm{NLO}+\mathrm{NNLL}^{H^{(2)}}}=158.7^{+5.6\,
  (3.6\%)}_{-6.9\,(4.3\%)} ~\mathrm{pb} \quad \mathrm{for} \quad H_{ij,{\bf
    I}}^{(2)}(1) = 0 
\\
&\quad& \sigma^{\mathrm{NLO}+\mathrm{NNLL}^{H^{(2)}}}=167.9^{+5.2\,
  (3.1\%)}_{-7.5\,(4.5\%)} ~\mathrm{pb} \quad \mathrm{for} \quad H_{ij,{\bf
    I}}^{(2)}(1) = \overline{H}_{ij,{\bf I}}^{(2)}(1) \; , 
\ea
where $A=2$ throughout.
These scale uncertainties are slightly larger than those found for
NNLO$_\beta$ in (\ref{eq:logON_1})-(\ref{eq:logON_4}), but they
are still small compared to the impact of the unknown finite
contributions of $H_{ij,{\bf I}}^{(2)}(1)$, as suggested by the
comparison between the $H=0$ and $H=\overline{H}$ results above. 

These remarks justify our choice not to include the $H_{ij,{\bf
    I}}^{(2)}$ function in our benchmark predictions for the central
value and the theoretical systematics.  We notice nevertheless that
the significant reduction in scale variation obtained with the
inclusion of the known, finite scale dependent terms, is indicative of
the uncertainty of the full NNLO result, when it will become
available.

\begin{figure}[t]
  \centering
     \hspace{0mm} 
   \includegraphics[width=0.48\textwidth]{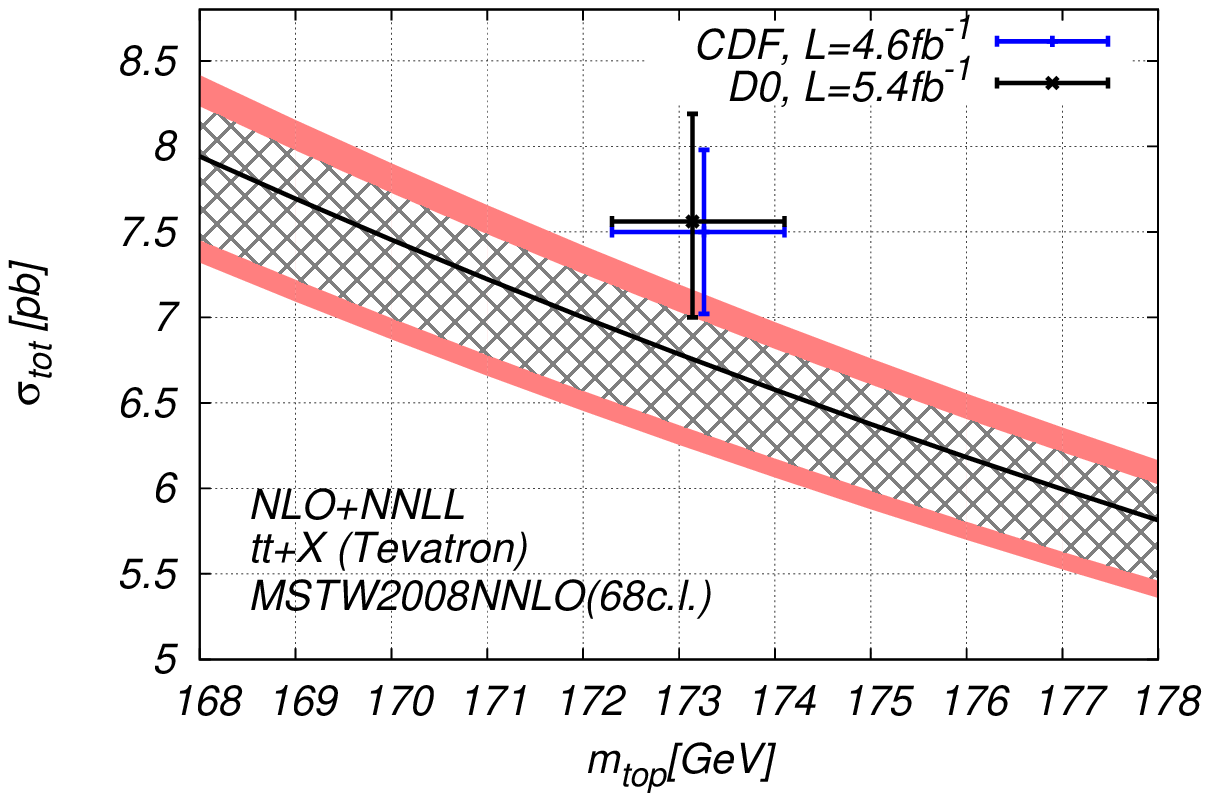} 
\hfill   \includegraphics[width=0.48\textwidth]{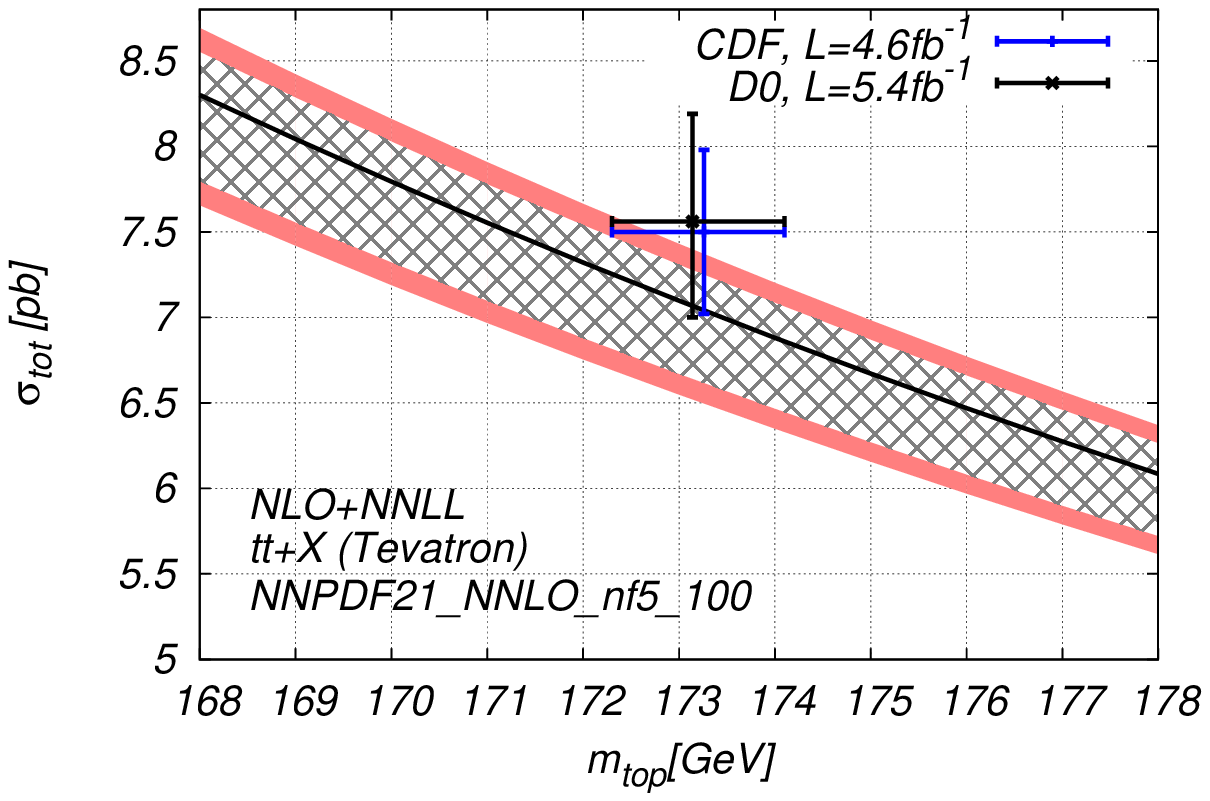} 
   \caption{ \label{fig:Tev-mass-range} Theoretical prediction for the
     total inclusive $t\bar t$ cross-section at the Tevatron, as a
     function of the top mass, versus the measurements of
     Refs.~\cite{Abazov:2011cq}, \cite{CDF-top-sigma}. The left plot
     is for the MSTW2008nnlo68cl PDF set, the right plot for 
     NNPDF21\_nnlo\_nf5\_100. The uncertainty
     is a linear sum of scale uncertainty (the white central band) and
     PDF uncertainty (red bands). The central value is shown with a
     black line. The theoretical predictions in this figure correspond
     to row 9 in table~\ref{tab:tev-syst}. The horizontal bars on the measurements reflect the uncertainty in the measured top mass (see footnote \ref{footnote} ). }
\end{figure}

\begin{figure}[t]
  \centering
     \hspace{0mm} 
   \includegraphics[width=0.48\textwidth]{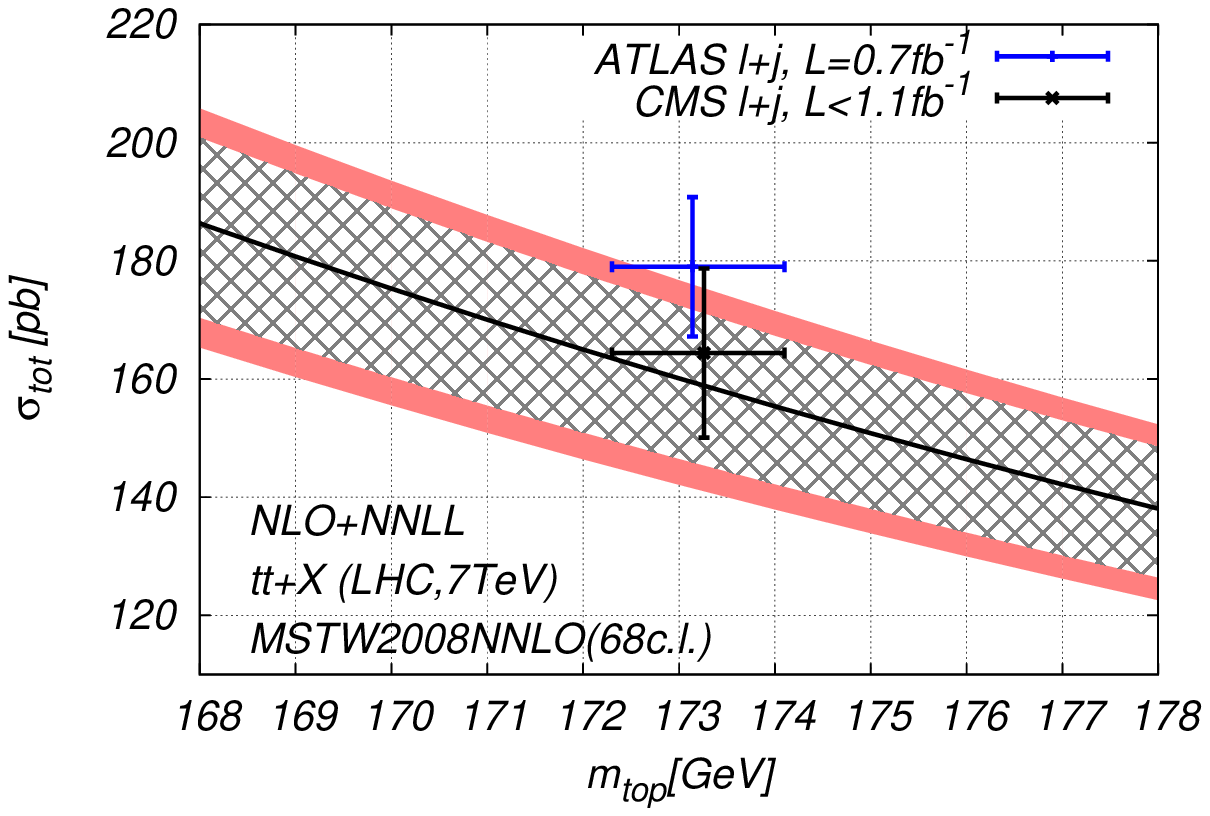} 
\hfill   \includegraphics[width=0.48\textwidth]{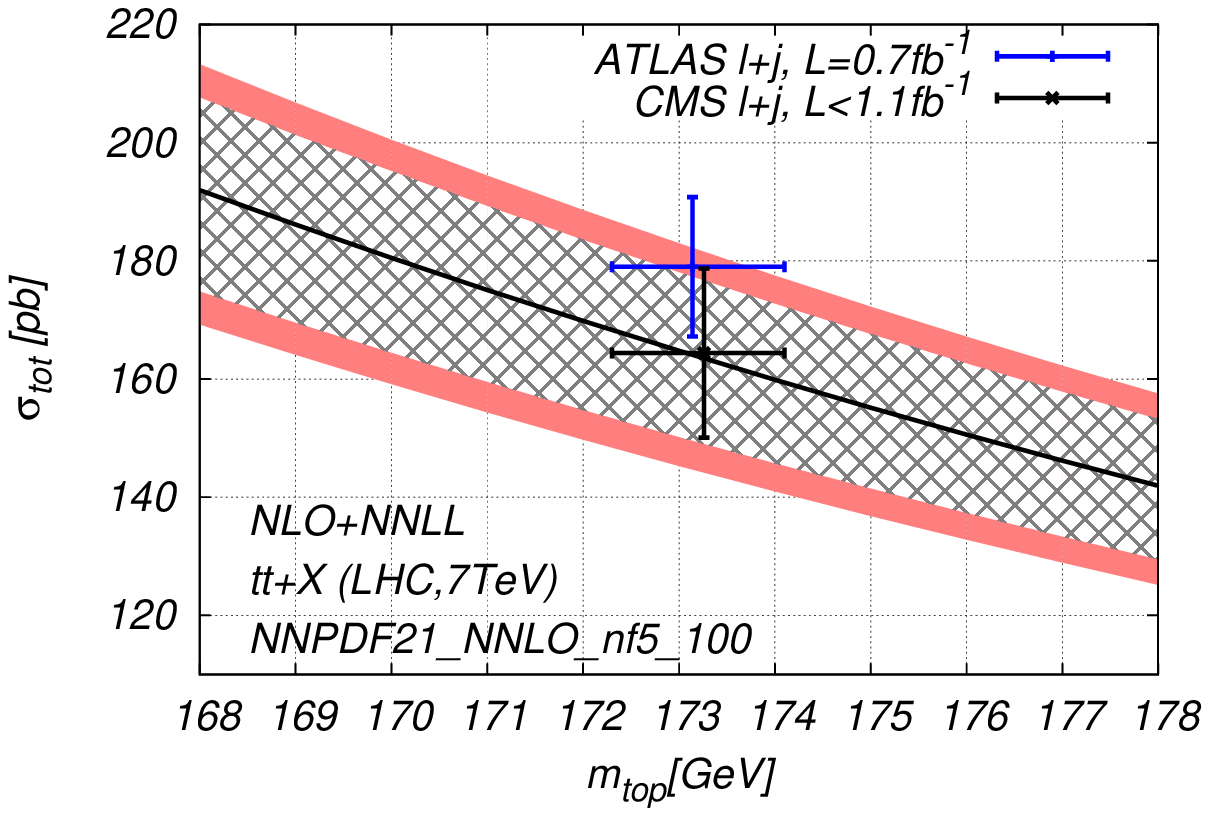} 
   \caption{\label{fig:LHC7-mass-range} Theoretical prediction for the
     total inclusive $t\bar t$ cross-section at the LHC @ 7 TeV as a
     function of the top mass versus the measurements of
     Refs.~\cite{ATLAS-top-sigma},\cite{CMS-top-sigma}. The left plot
     is for the MSTW2008nnlo68cl PDF set, the right plot for
     NNPDF21\_nnlo\_nf5\_100.  The uncertainty is a linear sum of scale
     uncertainty (the white central band) and PDF uncertainty (red
     bands). The central value is shown with a black line. The
     theoretical predictions in this figure correspond to row 9 in
     table~\ref{tab:lhc7-syst}. The horizontal bars on the measurements reflect the uncertainty in the measured top mass (see footnote \ref{footnote} ).}
\end{figure}

\section{Phenomenology}\label{sec:pheno}
It is well known that the purely theoretical uncertainty related to
the lack of high-order corrections is only a fraction of the overall
systematics. Recent studies of the $t\bar{t}$ cross section
uncertainty due to the PDF parameterization, including the latest fits
by several groups, have been reported
in~\cite{Watt:2011kp,Ball:2011uy}. These studies considered the
fixed-order NLO results and the approximate NNLO calculation of
Ref.~\cite{Moch:2008qy}, as implemented in the program
HATHOR~\cite{Aliev:2010zk}. The main conclusion of those studies was
the consistency, at the LHC energy, 
between the central values and the uncertainty bands
obtained using most PDF fits (MSTW08~\cite{Martin:2009iq},
NNPDF2.1~\cite{Ball:2011uy}, GJR~\cite{JimenezDelgado:2009tv},
CT10~\cite{Lai:2010vv}), both at NLO and NNLO, with some differences,
incompatible with the estimates of the systematics, with respect to
other sets such as ABKM09~\cite{Alekhin:2009ni} and
HERAPDF~\cite{:2009wt}.

Given the minor changes in fixed-order versus resummed predictions,
we expect that the PDF uncertainties estimated
in~\cite{Watt:2011kp,Ball:2011uy} should not be affected by
resummation. We verify this result explicitly here, considering 
our default PDF set
MSTW2008nnlo68cl~\cite{Martin:2009iq}, and the PDF set
NNPDF21\_nnlo\_nf5\_100~\cite{Ball:2011uy}.

Our results for the total
$t\bar t$ cross-section using MSTW2008nnlo68cl are:
\begin{eqnarray}
\sigma_{\rm tot}^{\mathrm{NLO}+\mathrm{NNLL}}({\rm Tevatron};~ m_t=173.3~{\rm GeV}) &=& 6.722^{~+0.238\, (3.5\%)}_{~-0.410\,(6.1\%)}~[{\rm scales}] 
^{~+0.160\,(2.4\%)}_{~-0.115\, (1.7\%)}~[{\rm PDF}] ~ {\rm pb}\, , \nonumber\\
&&\nonumber\\
\sigma_{\rm tot}^{\mathrm{NLO}+\mathrm{NNLL}}({\rm LHC}_{7{\rm TeV}};~m_t=173.3~{\rm GeV}) &=& 158.7^{~+12.2\, (7.7\%)}_{~-13.5\,(8.5\%)}~[{\rm scales}]
^{~+4.3 \,(2.7\%)}_{~-4.4\, (2.8\%)} ~[{\rm PDF}] ~ {\rm pb}\, ,
\label{eq:best-result}
\end{eqnarray}
where we defined the upper and lower limit of the scale variation
using the endpoint scan defined after Eq.~(\ref{eq:musyst}).
With 
NNPDF21\_nnlo\_nf5\_100
we obtain instead:
\footnote{As a central value we take the number derived with the central NNPDF set, not the mean over the whole set of PDFs. The difference is at the per mille level and thus completely negligible.}
\begin{eqnarray}
\sigma_{\rm tot}^{\mathrm{NLO}+\mathrm{NNLL}}({\rm Tevatron};~ m_t=173.3~{\rm GeV}) &=& 7.021^{~+0.250\,(3.6\%)}_{~- 0.436\,(6.2\%)}~[{\rm scales}] 
^{~+0.126\,(1.8\%)}_{~-0.119\, (1.7\%)}~[{\rm PDF}] ~ {\rm pb}\, , \nonumber\\
&&\nonumber\\
\sigma_{\rm tot}^{\mathrm{NLO}+\mathrm{NNLL}}({\rm LHC}_{7{\rm TeV}};~m_t=173.3~{\rm GeV}) &=& 163.1^{~+12.9\,(7.9\%)}_{~-14.2\,(8.7\%)}~[{\rm scales}]
^{~+4.9 \,(3.0\%)}_{~-4.9\, (3.0\%)} ~[{\rm PDF}] ~ {\rm pb}\, .
\label{eq:best-result-nnpdf}
\end{eqnarray}
The PDF uncertainties in both these sets of predictions should be
considered to be at the 1-$\sigma$ level.

A few comments are in order. To start with, we confirm that the relative
PDF uncertainty at NLO+NNLL is similar to that 
derived from a fixed order calculation, as
in~\cite{Watt:2011kp,Ball:2011uy}. We also notice that
the scale uncertainty is rather independent of the PDF set. This is
consistent with the fact that the relative contribution of $gg$,
$q\bar{q}$ and $qg$ initial states does not change significantly when
changing PDFs. 

We also confirm the consistency of the central value and of the PDF
uncertainty estimated, for the LHC at 7~TeV, using the MSTW and
the default NNPDF2.1 sets.  We note, on the other hand, that at the Tevatron the
NNPDF prediction is larger than MSTW by about $5\%$, compared to the individual
estimates of PDF systematics, which are of the order of $\pm 2\%$. 
This difference can be understood in terms of  the  different values of
the strong coupling constant associated with the MSTW2008nnlo68cl ($\as(M_Z) = 0.117$)
and the NNPDF21\_nnlo\_nf5\_100 ($\as(M_Z) = 0.119$) sets, and the fact that the 
cross section scales like $\as^2$. To better 
quantify this effect we have also used the set
NNPDF21\_nnlo\_as\_0117\_100~\cite{Ball:2011uy} to compute the central
values corresponding to Eq.~(\ref{eq:best-result-nnpdf}):
\begin{eqnarray}
\sigma_{\rm tot}^{\mathrm{NLO}+\mathrm{NNLL}}({\rm Tevatron};~ m_t=173.3~{\rm GeV}) &=& 6.742  ~ {\rm pb}\, , \nonumber\\
&&\nonumber\\
\sigma_{\rm tot}^{\mathrm{NLO}+\mathrm{NNLL}}({\rm LHC}_{7{\rm TeV}};~m_t=173.3~{\rm GeV}) &=& 156.8 ~ {\rm pb}\, .
\label{eq:best-result-nnpdf-as0117}
\end{eqnarray}
These results confirm that the apparent inconsistency between the Tevatron 
predictions of the NNPDF and MSTW NNLO sets
disappears when they both use the same coupling constant
($\alpha_S(M_Z)=0.117$ in this case). This also suggests that an additionally
uncertainty of the order of $\pm$~1-2$\%$ should likely be added to any $t\bar t$
total cross section evaluation as a result of the uncertainty with which $\as$ 
is known, if not already included with the PDF uncertainty.

We plot in Figs.~\ref{fig:Tev-mass-range} and
\ref{fig:LHC7-mass-range} our predictions for the total $t\bar t$
cross-section as a function of the top mass in the range $168-178~{\rm
  GeV}$ for both Tevatron and LHC ($\sqrt{S}=7$~TeV). 
 In view of the difference between the MSTW
and NNPDF results for the Tevatron, 
we present the results for the two PDF sets on different plots. 
The uncertainties from scales
and parton distributions quoted in Eq.~(\ref{eq:best-result}) are
added linearly. On the
same figures we compare our theoretical predictions with the most
accurate available experimental measurements from the Tevatron
\cite{Abazov:2011cq}, \cite{CDF-top-sigma} and LHC
\cite{ATLAS-top-sigma},\cite{CMS-top-sigma}.
We display the experimental
points at the current world-average value of $m_t=173.2\pm 0.9$~GeV,
without applying any correction factor to account for the difference
in experimental acceptance with respect to the $m_t$ values used in
the measurements. These amount to a reduction in rate at the sub-percent
level. 

We observe that the uncertainties of the current
theoretical predictions and experimental measurements  are comparable in size. 
The predictions agree
with all measurements within the uncertainties, although at the
Tevatron the data tend to be on high side of the theoretical band,
particularly in the case of the MSTW cross sections.

For ease of use, 
we have fitted the mass dependence of the 
$\sigma^{\mathrm{NLO}+\mathrm{NNLL}}$  predictions relative
to the MSTW sets using the functional form
\begin{equation}
\label{eq:parametrisation}
\sigma(m) = \sigma(m_{ref})\left(\frac{m_{ref}}{m}\right)^4
\left(1 + a_1\frac{m-m_{ref}}{m_{ref}} + a_2\left(\frac{m-m_{ref}}{m_{ref}}\right)^2\right) \, .
\end{equation}
The resulting parameters, for the central curve as well as the scales and PDF
uncertainties separately, are collected in table~\ref{table:params}. They provide fits that are 
accurate to within about one per mille in the 150-200 GeV mass range, but should not be used indiscriminately
much beyond this region.
\begin{table}[t]
\begin{center}
\begin{tabular}{|c|l|c|c|c|}
\hline
\multicolumn{2}{|c|}{${m_{ref}} = 173$ GeV} & $\sigma(m_{ref})$ (pb) & $a_1$  & $a_2$\\
\hline
\hline
\multirow{5}{*}{Tevatron, $p\bar p$ at $\sqrt{s}=1.96$ TeV }
& Central & 6.785 & $-$1.394 & 7.451 $\times 10^{-1}$    \\ 
& Scales $+$ & 7.030 & $-$1.409 & 8.047 $\times 10^{-1}$ \\ 
& Scales $-$ & 6.370 & $-$1.379 & 6.919 $\times 10^{-1}$ \\ 
& PDFs $+$ & 6.946 & $-$1.373 & 7.106 $\times 10^{-1}$   \\ 
& PDFs $-$ & 6.669 & $-$1.408 & 7.527 $\times 10^{-1}$   \\ 
\hline
\hline
\multirow{5}{*}{LHC, $pp$ at $\sqrt{s}=7$ TeV }
& Central & 160.1 & $-$1.191 & 8.042 $\times 10^{-1}$     \\ 
& Scales $+$ & 172.4 & $-$1.224 & 9.096 $\times 10^{-1}$  \\ 
& Scales $-$ & 146.5 & $-$1.162 & 7.957 $\times 10^{-1}$  \\ 
& PDFs $+$ & 164.4 & $-$1.175 & 7.867 $\times 10^{-1}$    \\ 
& PDFs $-$ & 155.7 & $-$1.205 & 8.416 $\times 10^{-1}$    \\ 
\hline
\hline
\multirow{5}{*}{LHC, $pp$ at $\sqrt{s}=8$ TeV }
& Central & 228.6 & -1.069 & 6.798 $\times 10^{-1}$  \\
& Scales $+$ & 246.8 & -1.104 & 7.335 $\times 10^{-1}$ \\
& Scales $-$ & 208.8 & -1.042 & 6.299 $\times 10^{-1}$ \\
& PDFs $+$ & 234.2 & -1.054 & 6.533 $\times 10^{-1}$  \\
& PDFs $-$ & 222.7 & -1.083 & 6.964 $\times 10^{-1}$  \\
\hline
\end{tabular}
\caption{\label{table:params} Parameters resulting from the fit of the functional
form in eq.~(\ref{eq:parametrisation}) to our best prediction, NLO+NNLL, for the top cross section
at the Tevatron and the LHC (7 and 8 TeV) as a function of the top mass, 
including the uncertainties from scale variations and PDFs (the MSTW2008nnlo68cl set).
These parameters provide fits that are 
accurate to within about one per mille in the 150-200 GeV mass range, but should not be used indiscriminately
much beyond this region.}
\end{center}
\end{table}

\begin{figure}[t]
  \centering
     \hspace{0mm} 
   \includegraphics[width=11cm]{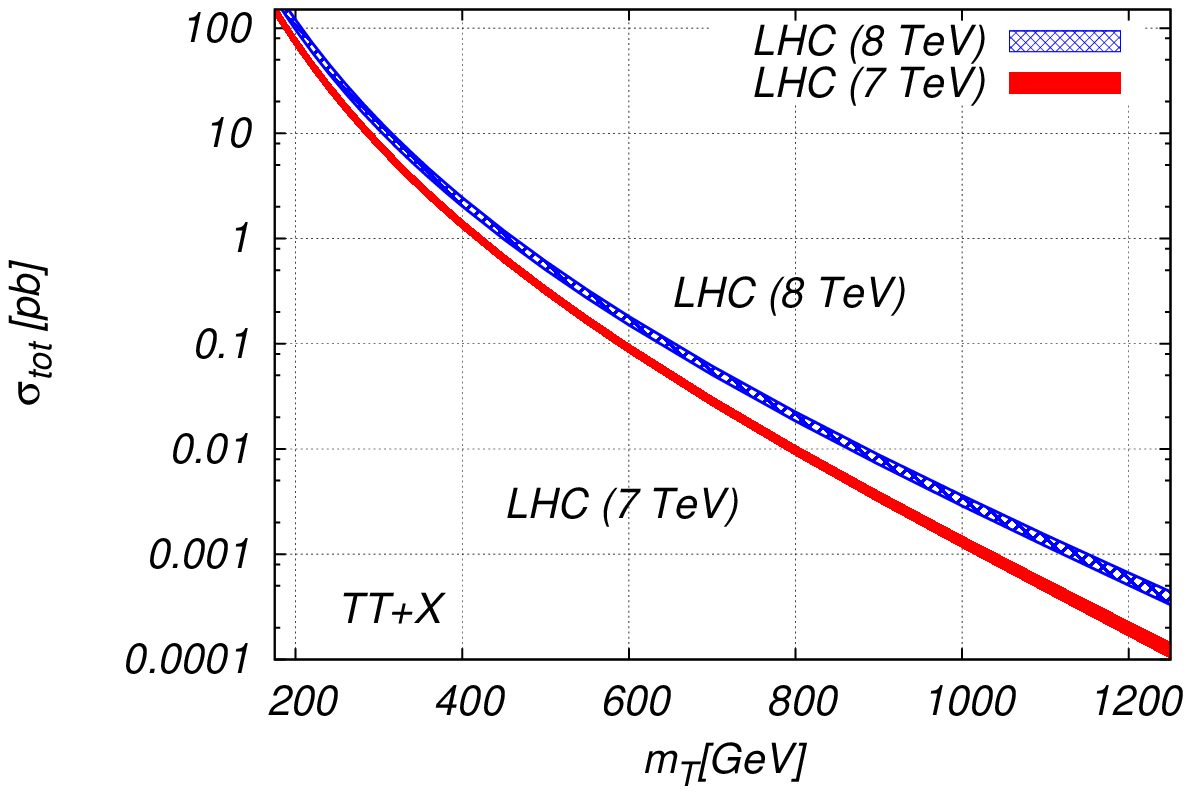} 
   \caption{Theoretical prediction for the total inclusive production
     cross-section at the LHC (7 and 8 TeV) of a $T\bar T$ 
   pair of hypothetical heavy fourth-generation quarks. Effects from the running of top quarks in loops
   are expected to be very small and are neglected in these numerical predictions.}
      \label{fig:LHC7-heavy-mass}
\end{figure}

Finally, due to the interest in current experimental searches of
possible~\cite{Holdom:2009rf} fourth-generation quarks, both at the
Tevatron~\cite{Aaltonen:2011vr,Aaltonen:2011tq,Abazov:2011vy} and at
the LHC~\cite{Chatrchyan:2011em}, we extend our results to the
production of a hypothetical very heavy quark, $T$. Its production
cross section at the LHC (for $\sqrt{S}=7$ and 8~TeV) 
up to $m_T$ = 1200 GeV is shown in
figure~\ref{fig:LHC7-heavy-mass}, and the detailed breakdown of the
systematics is given, for a set of mass values, in
Table~\ref{tab:highm}. The cross sections are calculated by simply
changing the mass parameter value in the top cross section
calculation, thus neglecting the small corrections due to the top
quark in the evolution of $\alpha_s$ and of PDFs.
 
The overall uncertainty (scale+PDF) is roughly uniform at the
$\pm$10-15\% level in the range of the plot. This is the result of a
decreasing scale uncertainty, which is more than compensated by an
increasing PDF uncertainty, due to the large partonic $x$ values
probed by the production of a large mass object, and PDFs being
generally less well known in this region.

Finally, we present our best predictions for the LHC
configuration foreseen for the 2012 data taking ($\sqrt{S}=$ 8 TeV), 
with the PDF sets MSTW2008nnlo68cl~\cite{Martin:2009iq}:
\begin{eqnarray}
\sigma_{\rm tot}^{\mathrm{NLO}+\mathrm{NNLL}}({\rm LHC}_{8{\rm TeV}};~m_t=173.3~{\rm GeV}) &=& 226.6^{~+17.8\,(7.8\%)}_{~-19.4\,(8.6\%)}~[{\rm scales}]
^{~+5.6 \,(2.5\%)}_{~-5.8\, (2.6\%)} ~[{\rm PDF}] ~ {\rm pb}\, .
\label{eq:best-result-mstw-lhc8}
\end{eqnarray}
and NNPDF21\_nnlo\_nf5\_100~\cite{Ball:2011uy}:
\begin{eqnarray}
\sigma_{\rm tot}^{\mathrm{NLO}+\mathrm{NNLL}}({\rm LHC}_{8{\rm TeV}};~m_t=173.3~{\rm GeV}) &=& 233.5^{~+18.9\,(8.1\%)}_{~-20.5\,(8.8\%)}~[{\rm scales}]
^{~+6.5 \,(2.8\%)}_{~-6.5\, (2.8\%)} ~[{\rm PDF}] ~ {\rm pb}\, .
\label{eq:best-result-nnpdf-lhc8}
\end{eqnarray}

All of the numerical results presented in this section, and their
extrapolation to different values of the heavy quark mass or of the
scale parameters, can be obtained through a simple web
interface~\cite{web-interface}. These, and more general results
obtained under the various approximation scenarios outlined in
Section~\ref{sec:syst}, can furthermore be computed with the help of
the program {\tt Top++}~\cite{Czakon:2011xx}.

\begin{table}[ht]
\begin{center}
\begin{tabular}{| c | c | c | c |c | c | c |}
\hline
 & 
\multicolumn{3}{|c|}{LHC, $\sqrt{S}=7$~TeV} &
\multicolumn{3}{|c|}{LHC, $\sqrt{S}=8$~TeV} \\

\hline
$m_T$ (GeV)  
& $\sigma_{\rm tot}~[{\rm pb}]$ 
&  Scale 
& PDF  
& $\sigma_{\rm tot}~[{\rm pb}]$ 
&  Scale 
& PDF  \\
\hline
 200 &    74.71 & $^{+   5.498\, (  7.4\%)}_{-   6.165\, (  8.3\%)}$ &
 $^{+   2.189\, (  2.9\%)}_{-   2.205\, (  3.0\%)}$
& 108.59 & $_{   -9.022 (8.3\%) }^{+   8.100 (7.5\%) } $ & $_{   -2.980 (2.7\%)
                          }^{+2.940 (2.7\%) } $
\\ \hline
 300 &     7.83 & $^{+   0.469\, (  6.0\%)}_{-   0.563\, (  7.2\%)}$ &
 $^{+   0.275\, (  3.5\%)}_{-   0.272\, (  3.5\%)}$
& 12.09  & $_{   -0.882 (7.3\%) }^{+   0.750 (6.2\%) } $ & $_{   -0.397 (3.3\%)
                        }^{+0.399 (3.3\%) } $
\\ \hline
 400 &     1.38 & $^{+   0.070\, (  5.1\%)}_{-   0.089\, (  6.5\%)}$ &
 $^{+   0.053\, (  3.9\%)}_{-   0.052\, (  3.8\%)}$
& 2.25   & $_{   -0.148 (6.6\%) }^{+   0.119 (5.3\%) } $ & $_{   -0.081 (3.6\%)
                        }^{+0.082 (3.7\%) } $
\\ \hline
 500 &     0.32 & $^{+   0.015\, (  4.6\%)}_{-   0.019\, (  6.0\%)}$ &
 $^{+   0.014\, (  4.3\%)}_{-   0.013\, (  4.1\%)}$
&  0.56   & $_{   -0.034 (6.1\%) }^{+   0.026 (4.6\%) } $ & $_{   -0.022 (3.9\%)
                         }^{+0.022 (4.0\%) } $
\\ \hline

$m_T$ (GeV)  
& $\sigma_{\rm tot}~[{\rm fb}]$ 
&  Scale 
& PDF 
& $\sigma_{\rm tot}~[{\rm fb}]$ 
&  Scale 
& PDF  \\
\hline

 600 &    90.80 & $^{+   3.945\, (  4.3\%)}_{-   5.059\, (  5.6\%)}$ &
 $^{+   4.380\, (  4.8\%)}_{-   4.031\, (  4.4\%)}$
& 166.74 & $_{   -9.480 (5.7\%) }^{+   7.290 (4.4\%) } $ & $_{   -6.860 (4.1\%)
                        }^{+  7.240 (4.3\%) }$ 
\\ \hline
 700 &    28.60 & $^{+   1.191\, (  4.2\%)}_{-   1.512\, (  5.3\%)}$ &
 $^{+   1.613\, (  5.6\%)}_{-   1.406\, (  4.9\%)}$
& 56.04  & $_{   -3.021 (5.4\%) }^{+   2.349 (4.2\%) } $ & $_{   -2.475 (4.4\%)
                        }^{+  2.747 (4.9\%) }$ 
\\ \hline
 800 &     9.76 & $^{+   0.391\, (  4.0\%)}_{-   0.494\, (  5.1\%)}$ &
 $^{+   0.652\, (  6.7\%)}_{-   0.546\, (  5.6\%)}$
& 20.50  & $_{   -1.058 (5.2\%) }^{+   0.827 (4.0\%) } $ & $_{   -0.995 (4.9\%)
                        }^{+  1.153 (5.6\%) } $
\\ \hline
 900 &     3.52 & $^{+   0.135\, (  3.8\%)}_{-   0.172\, (  4.9\%)}$ &
 $^{+   0.280\, (  8.0\%)}_{-   0.229\, (  6.5\%)}$
& 7.97   & $_{   -0.397 (5.0\%) }^{+   0.310 (3.9\%) } $ & $_{   -0.436 (5.5\%)
                        }^{+  0.518 (6.5\%) } $
\\ \hline
1000 &     1.32 & $^{+   0.049\, (  3.7\%)}_{-   0.063\, (  4.8\%)}$ &
$^{+   0.125\, (  9.5\%)}_{-   0.101\, (  7.7\%)}$
& 3.24   & $_{   -0.157 (4.8\%) }^{+   0.122 (3.8\%) } $ & $_{   -0.202 (6.2\%)
                        }^{+	0.246 (7.6\%) } $
\\ \hline
1100 &     0.51 & $^{+   0.018\, (  3.6\%)}_{-   0.023\, (  4.6\%)}$ &
$^{+   0.057\, ( 11.2\%)}_{-   0.046\, (  9.1\%)}$
& 1.36   & $_{   -0.064 (4.7\%) }^{+   0.049 (3.6\%) } $ & $_{   -0.098 (7.2\%)
                        }^{+	0.120 (8.9\%) } $
\\ \hline
1200 &     0.20 & $^{+   0.007\, (  3.5\%)}_{-   0.009\, (  4.6\%)}$ &
$^{+   0.026\, ( 13.1\%)}_{-   0.021\, ( 10.7\%)}$
& 0.58   & $_{   -0.027 (4.6\%) }^{+   0.021 (3.5\%) } $ & $_{   -0.049 (8.3\%)
                        }^{+  0.060 (10.3\%) } $
\\ \hline

\end{tabular}
\caption{\label{tab:highm} Total cross sections, at NLO+NNLL level,
  for the production of a heavy quark at the LHC ($\sqrt{S}=7$ and
  8~TeV), including the uncertainties from scale variations and PDFs
  (using the MSTW2008nnlo68cl set). }
\end{center}
\end{table}


\section{Concluding remarks}\label{sec:conclusions}
Using recent theoretical developments, we extend in this paper the
soft-gluon resummation of the total $t\bar t$ cross-section to the
NNLL order, using the Mellin $N$-space formalism. The result includes all
known NNLO terms that are singular at the production threshold.
The current work represents the third-order logarithmic improvement
for this important collider observable that has been instrumental in
developments in precision collider physics.

We  explored the implications of the NNLL approximation in a
comprehensive phenomenological study of the total $t\bar t$
cross-section, to quantify the full theoretical uncertainty currently
associated with this observable.

In fixed order calculations the theoretical uncertainty is typically
identified with the residual scale sensitivity of an observable. While
not perfect, such a procedure is well understood and gives a
meaningful way of comparing theoretical predictions across different
observables and levels of precision. The procedure relies on the
following considerations: since the exact result must be
scale-independent, and the scale dependence of each fully calculated order
must be of higher order, 
one assumes that the residual scale dependence of a calculation is numerically comparable to the higher-order
scale-independent terms, which can only be obtained via the complete
calculation, and whose size the theoretical systematics attempt to
estimate\footnote{A detailed presentation of these well-known assumptions 
is given in \cite{Cacciari:2011ze}, where they are then used to argue for
an alternative way of characterizing the perturbative theoretical uncertainty.
The method proposed in this paper has however so far only been detailed for
$e^+e^-$ collisions, and cannot therefore be applied to hadronic top
production.}. 
When dealing with an approximate NNLO calculation for $t\bar t$ hadroproduction, and trying to assess its uncertainty via the residual scale dependence, one must keep in mind that
known terms of ${\cal O}(\as^4)$
include: (a) terms singular at the production threshold, both scale
dependent and independent, whose behaviour at higher orders is
determined by general dynamical considerations, and which can
therefore be included and resummed, with a genuine improvement of the
accuracy; (b) finite, but scale-dependent, terms, whose value can be
fixed by imposing full ${\cal O}(\as^4)$ scale-independence. Inclusion of such terms will lead, by construction, to a reduction of the scale
dependence, but this reduction does not reflect the real size of the
theoretical uncertainty, which is rather governed by the unknown constant
terms of ${\cal O}(\as^4)$.
   
In this paper we assessed the possible size of several unknown
higher-order contributions, and studied their contribution to the
theoretical uncertainty. In particular, we demonstrated that the
reduced scale sensitivity, obtained by using the exact ${\cal O}(\as^4)$
scale dependence, leads, once the uncertainty of the unknown terms is
accounted for, to a larger overall systematics, comparable to that of
the NLO+NLL cross section. 

We also demonstrated that the predicted cross section has a
few-percent sensitivity to currently unknown $1/N$ suppressed terms
that are beyond any of the approximations available in the
literature. Summarizing these observations we conclude that at present
the total uncertainty of the total $t\bar t$ cross-section at the
NLO+NNLL order is only modestly lower compared to the long-established
NLO+NLL result. Guided by the small scale dependence of the results
obtained imposing the exact ${\cal O}(\as^4)$ scale dependence, we
nevertheless speculate a significant decrease of the theoretical
uncertainty in the total $t\bar t$ cross-section once the full NNLO
result becomes available.

Finally we would like to briefly compare our work with theoretical
works that have appeared in the recent past and that make, to various
extent, use of NNLO approximations.

Reference~\cite{Moch:2008qy} uses a fixed-order approximate NNLO 
approach to the total
inclusive cross-section, including the 
$C^{(2,n)}$ terms that implement the exact ${\cal O}(\as^4)$ scale
dependence. The overall uncertainty is estimated by just varying the scale
in this framework, without accounting for the uncertainty of the
finite $C^{(2,0)}$ pieces, leading, as we argued above, to a much reduced and
in our view optimistic systematics.
This calculation has been implemented in the program HATHOR~\cite{Aliev:2010zk}.

Reference~\cite{Beneke:2011mq} pursues a resummation approach
that shares many similarities with our work. Its authors resum directly the total
inclusive cross-section by implementing the same anomalous dimensions
and 2-loop Coulomb terms used here, and do not impose exact 
${\cal O}(\as^4)$ scale
dependence.
The resummation method instead differs. In Ref.~\cite{Beneke:2011mq}
the so called momentum space approach of Ref.~\cite{Becher:2006nr} is
used, which is an $x$-space approach, while we use an $N$-space
resummation, followed by a Mellin inversion. In the approach of
Ref.~\cite{Becher:2006nr}, the $x$-space perturbative expansion of the
resummed cross section is convergent, while in our approach the
perturbative expansion of our $N$-space result is convergent, and its
Mellin inversion to $x$ space is asymptotic. This feature has been
criticized as a drawback of the Mellin space approach\footnote{ In
  ref.~\cite{Becher:2006nr} it is claimed that integration over the
  Landau pole also arise in the computation of the $N$-space
  resummation formula. We remark, however, that this pole is
  irrelevant for the derivation of the N space formula, that in fact
  has a convergent perturbative expansion.} in
Ref.~\cite{Becher:2006nr}. However, we remind the reader that the
ambiguity associated with the asymptotic nature of the Mellin
inversion is very weak, corresponding to an effect that is suppressed
more strongly than any inverse power of the process scale, and that in
practice has totally negligible effects. Furthermore, factorization in
$N$ guarantees naturally momentum conservation. Although momentum
conservation can be abandoned in the soft approximation, it was shown
in~\cite{Catani:1996dj} that it can lead to large subleading
effects. We thus believe that the Mellin space approach is worth
pursuing for this positive feature\footnote{A critical comparison of
  the $x$- and $N$-space method at the analytic level has been presented
  in ref.~\cite{Bonvini:2012yg} for the case of Drell-Yan pair
  production.}.

At the approximate NNLO order, our NNLO$_\beta$ rates and scale
systematics for $C^{(2,0)}=0$ (Tables~\ref{tab:tev-syst} and
\ref{tab:lhc7-syst}) agree precisely with the equivalent results,
labeled NNLO$_{\mathrm{app}}$, in Tables 8 and 10
of~\cite{Beneke:2011mq}.  After resummation, the differences with our
work in the final predictions and theoretical systematics must be
attributed to the different formalism. One such source of difference,
for example, is that in the $x$-space resummation approach additional
scales are present (in our case these are only $\mu_F$ and
$\mu_R$). We also note the different default values used for the
unknown two-loop constants, which is also a reflection of the
different formalisms used, since in the $N$-space approach the
resummed terms vanish for $N=1$ by construction.\footnote{This is to
  avoid introducing corrections at $N=1$ from a formalism that is
  valid at large $N$.}  It is perhaps surprising that the largest
difference among central values is observed for the Tevatron, while at
the LHC central values are very close. This could be related to the
observation made in~\cite{Beneke:2011mq}, namely that the contribution
of the $q\bar{q}$ channel is poorly approximated by the threshold
expansion. Due to the dominance of this channel at the Tevatron,
Ref.~\cite{Beneke:2011mq} argues that this could also explain why
the Tevatron uncertainty does not improve after NNLL resummation. 
In all cases, the numerical differences are nevertheless
consistent with the overall uncertainties quoted both in our work and
in~\cite{Beneke:2011mq}.

An alternative approach to the total $t \bar t$ cross-section has been
pursued in
Refs.~\cite{Ahrens:2010zv,Ahrens:2011mw,Ahrens:2011px}. Like
Ref.~\cite{Beneke:2011mq},
Refs.~\cite{Ahrens:2010zv,Ahrens:2011mw,Ahrens:2011px} are based on the
momentum space approach. There are a number of additional differences
between our work and these papers. They resum not the total but the
differential cross-section for $t\bar t$ production. Once resummation
is performed at the differential level, the differential distribution
is integrated over phase space to obtain the total inclusive rate. In
such an approach the leading terms at absolute threshold are correctly
reproduced (see Ref.~\cite{Czakon:2009zw}) but one introduces a
different set of unknown power corrections (as also pointed out in
Ref.~\cite{Beneke:2011mq}). Various choices for the hard scales, which
become available when calculating differential quantities, have been
explored in
Refs.~\cite{Ahrens:2010zv,Ahrens:2011mw,Ahrens:2011px}. While the
central value of the total cross section in Ref.~\cite{Ahrens:2011px}
is slightly different from ours and from the results of
Ref.~\cite{Beneke:2011mq}, it is reassuring that they all still fit
within the quoted uncertainty bands (the minor difference in the
reference top mass in~\cite{Ahrens:2011px}, $m_{top}=173.1$~GeV, has
no impact in this comparison).

Overall, the non-PDF related uncertainties in the $x$-space
resummation approaches \cite{Ahrens:2011px,Beneke:2011mq} tend to be
smaller than ours (by between about $25\%$ to $40\%$), with the
exception of the Tevatron prediction of Ref.~\cite{Beneke:2011mq},
which has a corresponding uncertainty about $20\%$ larger than ours.

An approach that shares similarities with
\cite{Ahrens:2010zv,Ahrens:2011mw,Ahrens:2011px} has been pursued by
Ref.~\cite{Kidonakis:2010dk}. In that reference an approximate, fixed
order truncation of the differential cross section is derived and
 integrated over phase space to obtain the fully inclusive
cross section. For the LHC the scale variation and central values
derived in Ref.~\cite{Kidonakis:2010dk} are similar to those of
Ref.~\cite{Aliev:2010zk} (which is about $50\%$ smaller than our
benchmark result). For the Tevatron the central values of
these two references are also rather close, while the uncertainty of
Ref.~\cite{Kidonakis:2010dk} is much smaller than that of all other
groups (it is about $60\%$ smaller than ours); 
see also Refs.~\cite{Kidonakis:2011ca}.

Similarities and differences between some of the approaches above have
already been addressed in
Refs.~\cite{Kidonakis:2011ca,Beneke:2011mq}. Such significant
differences can be partially understood with the help of the
discussion in Section~\ref{sec:syst}. Overall, the large differences
between central values and systematics reported in the various papers
discussed in this section appear to be another confirmation of our
conclusions about the precision with which the total $t \bar t$
cross-section is presently calculated.

\noindent
\section*{Acknowledgments}
We thank M. Beneke, P. Falgari, S. Klein, S. Moch, C. Schwinn and P. Uwer
for discussions and J. Rojo for clarifications about the NNPDF sets used in this work. 
M.Ca. was supported in part by grant ANR-09-BLAN-0060 of the French
Agence National de la Recherche, and by the EU ITN grant LHCPhenoNet, PITN-GA-2010-264564. 
M.Cz. was supported by the Heisenberg and by the Gottfried  Wilhelm Leibniz Programmes of the 
Deutsche Forschungsgemeinschaft. The work of A.M. was supported in part by the U.S. National 
Science Foundation, grant NSF-PHY-0705682, the LHC Theory Initiative, Jonathan Bagger, PI.

\appendix
\section{Properties of the finite threshold terms}
\label{app:nnlo-approx}
The presently unknown constants $C^{(2,0)}_{ij,\mathbf{I}}$ introduced
in Eq.~(\ref{eq:nnlo}) are related to the also unknown constants
$H_{ij,{\bf I}}^{(2)}(1)$ appearing in the two-loop hard matching
function (\ref{eq:Hard}). With a direct calculation, and presenting
directly numerical values, we obtain:
\begin{eqnarray}
C^{(2,0)}_{q{\bar q},\mathbf{8}} &=&  -489.168 + 16\, H_{q{\bar q},{\bf 8}}^{(2)}(1) \, , \nonumber\\ 
C^{(2,0)}_{gg,\mathbf{8}} &=& -1334.18 + 16\, H_{gg,{\bf 8}}^{(2)}(1)  \, , \nonumber\\
C^{(2,0)}_{gg,\mathbf{1}} &=&  -643.397  + 16\, H_{gg,{\bf 1}}^{(2)}(1)  \, ,
\label{eq:CtoH}
\end{eqnarray}
which, for the case $H=0$, results in the following combinations of
the constants $C^{(2,0)}_{ij,\mathbf{I}}$ that enter the
color-averaged cross-section:
\begin{eqnarray}
\overline{C}^{(2,0)}_{gg} &=& -1136.81 \, , \nonumber\\
\overline{C}^{(2,0)}_{q{\bar q}} &=&  -489.168 \, .
\label{eq:Cbar}
\end{eqnarray}
In an analogous way we shall define as $\overline{H}^{(2)}(1)$ the
values of the ${H}^{(2)}(1)$ constants obtained when setting 
$C^{(2,0)}_{ij,\mathbf{I}}=0$. 
We note that the constants
$C^{(2,0)}_{ij,\mathbf{I}}$ and $H_{ij,{\bf I}}^{(2)}(1)$ are defined
in different normalizations ($\as/(4\pi)$ in
Eq.~(\ref{eq:sigma2-color}) and $\as/\pi$ in Eq.~(\ref{eq:Hard})).

The dependence of the cross-section on the constants
$C^{(2,0)}_{ij,\mathbf{I}}$ can be estimated from:
\be
\label{eq:Csyst}
\sigma_{\rm tot} = 
\sigma_{\rm tot}(C^{(2,0)}_{ij,\mathbf{I}} =0) + 
\left({C^{(2,0)}_{qq,\mathbf{8}}\over 1000}\right) \Delta_{qq,\mathbf{8}} + 
\left({C^{(2,0)}_{gg,\mathbf{1}}\over 1000} \right) \Delta_{gg,\mathbf{1}} + 
\left({C^{(2,0)}_{gg,\mathbf{8}}\over 1000} \right) \Delta_{gg,\mathbf{8}} 
\, .
\ee
For $\mu_R=\mu_F=m=173.3\,$GeV \cite{:1900yx}, and with PDF set MSTW2008nnlo68cl~\cite{Martin:2009iq}, the values of $\Delta_{ij,\mathbf{I}}$ are provided in Table~\ref{tab:Delta}.
\begin{table}[t]
\caption{Values of  $\Delta_{ij,\mathbf{I}}$ in ${\rm pb}$ for 
$\mu_R=\mu_F=m=173.3\,$GeV and the MSTW2008nnlo68cl PDF set.}
\vskip0.2cm
\centering
\begin{tabular}{|l|ccc|}
\hline
\mbox{Collider} &   $\Delta_{qq,\mathbf{8}}$ & 
$\Delta_{gg,\mathbf{8}}$ & $\Delta_{gg,\mathbf{1}}$ \\
\hline
\mbox{Tevatron} & 0.3452 & 0.0241 & 0.0079 \\ \hline
\mbox{LHC7} & 1.698 & 4.313 & 1.305 \\ \hline
\mbox{LHC14} & 5.338 & 27.14 & 7.967 \\ \hline
\end{tabular}
\label{tab:Delta}
\end{table}
Clearly, reasonable variation of the unknown constants results in
variation of the predicted cross-section by a few percent, setting an
intrinsic limit to the precision of any estimate in absence of the
full knowledge of the NNLO result.

\end{document}